\documentclass[longauth]{aa}
\usepackage{natbib}
\usepackage[varg]{txfonts}
\usepackage{graphicx}

\usepackage{booktabs}
\usepackage{multirow}
\usepackage{tabularx}
\usepackage{lscape}
\usepackage{longtable}

\newcolumntype{L}[1]{>{\raggedright\arraybackslash}p{#1}}
\newcolumntype{C}[1]{>{\centering\arraybackslash}p{#1}}
\newcolumntype{R}[1]{>{\raggedleft\arraybackslash}p{#1}}

\usepackage{color} 

\def\Msun{\ifmmode{\mathrm M_\odot}\else{$M_\odot$}\fi}

\def\mod#1{{#1}}

\begin{document}

\title{Azimuthal offsets in spiral arms of nearby galaxies}


\newcommand{\OSU}{\label{OSU} Department of Astronomy, The Ohio State University, 140 West 18th Avenue, Columbus, Ohio 43210, USA}

\newcommand{\Alberta}{\label{Alberta} Department of Physics, University of Alberta, Edmonton, AB T6G 2E1, Canada}

\newcommand{\ANU}{\label{ANU} Research School of Astronomy and Astrophysics, Australian National University, Canberra, ACT 2611, Australia}

\newcommand{\IPAC}{\label{IPAC} Caltech-IPAC, 1200 E. California Blvd. Pasadena, CA 91125, USA}

\newcommand{\Carnegie}{\label{Carnegi} Observatories of the Carnegie Institution for Science, 813 Santa Barbara Street, Pasadena, CA 91101, USA}

\newcommand{\CCAPP}{\label{CCAPP} Center for Cosmology and Astroparticle Physics, 191 West Woodruff Avenue, Columbus, OH 43210, USA}

\newcommand{\CfA}{\label{CfA} Harvard-Smithsonian Center for Astrophysics, 60 Garden Street, Cambridge, MA 02138, USA}

\newcommand{\CITEVA}{\label{CITEVA} Centro de Astronomía (CITEVA), Universidad de Antofagasta, Avenida Angamos 601, Antofagasta, Chile}

\newcommand{\CNRS}{\label{CNRS} CNRS, IRAP, 9 Av. du Colonel Roche, BP 44346, F-31028 Toulouse cedex 4, France}

\newcommand{\ESO}{\label{ESO} European Southern Observatory, Karl-Schwarzschild Stra{\ss}e 2, D-85748 Garching bei M\"{u}nchen, Germany}

\newcommand{\Heidelberg}{\label{Heidelberg} Astronomisches Rechen-Institut, Zentrum f\"{u}r Astronomie der Universit\"{a}t Heidelberg, M\"{o}nchhofstra\ss e 12-14, D-69120 Heidelberg, Germany}

\newcommand{\COOL}{\label{COOL} Cosmic Origins Of Life (COOL) Research DAO, coolresearch.io}

\newcommand{\ICRAR}{\label{ICRAR} International Centre for Radio Astronomy Research, University of Western Australia, 35 Stirling Highway, Crawley, WA 6009, Australia}

\newcommand{\IRAM}{\label{IRAM} Institut de Radioastronomie Millim\'{e}trique (IRAM), 300 Rue de la Piscine, F-38406 Saint Martin d'H\`{e}res, France}

\newcommand{\IRAP}{\label{IRAP} IRAP, OMP, Université de Toulouse, 9 Av. du Colonel Roche, BP 44346, F-31028 Toulouse cedex 4, France}

\newcommand{\ITA}{\label{ITA} Universit\"{a}t Heidelberg, Zentrum f\"{u}r Astronomie, Institut f\"{u}r Theoretische Astrophysik, Albert-Ueberle-Str 2, D-69120 Heidelberg, Germany}

\newcommand{\IWR}{\label{IWR} Universit\"{a}t Heidelberg, Interdisziplin\"{a}res Zentrum f\"{u}r Wissenschaftliches Rechnen, Im Neuenheimer Feld 205, D-69120 Heidelberg, Germany}

\newcommand{\JHU}{\label{JHU} Department of Physics and Astronomy, The Johns Hopkins University, Baltimore, MD 21218, USA}

\newcommand{\Leiden}{\label{Leiden} Leiden Observatory, Leiden University, P.O. Box 9513, 2300 RA Leiden, The Netherlands}

\newcommand{\Maryland}{\label{Maryland} Department of Astronomy, University of Maryland, College Park, MD 20742, USA}

\newcommand{\MPE}{\label{MPE} Max-Planck-Institut f\"{u}r extraterrestrische Physik, Giessenbachstra{\ss}e 1, D-85748 Garching, Germany}

\newcommand{\MPIA}{\label{MPIA} Max-Planck-Institut f\"{u}r Astronomie, K\"{o}nigstuhl 17, D-69117, Heidelberg, Germany}

\newcommand{\Nagoya}{\label{Nagoya} Department of Physics, Nagoya University, Furo-cho, Chikusa-ku, Nagoya, Aichi 464-8602, Japan}

\newcommand{\NRAO}{\label{NRAO} National Radio Astronomy Observatory, 520 Edgemont Road, Charlottesville, VA 22903-2475, USA}

\newcommand{\NRAOAb}{\label{NRAOAb} National Radio Astronomy Observatory, 800 Bradbury SE, Suite 235, Albuquerque, NM 87106, USA}

\newcommand{\OAN}{\label{OAN} Observatorio Astron\'{o}mico Nacional (IGN), C/Alfonso XII, 3, E-28014 Madrid, Spain}

\newcommand{\ObsParis}{\label{ObsParis} Sorbonne Universit\'{e}, Observatoire de Paris, Universit\'{e} PSL, CNRS, LERMA, F-75014, Paris, France}

\newcommand{\Princeton}{\label{Princeton} Department of Astrophysical Sciences, Princeton University, 4 Ivy Ln., Princeton, NJ 08544 USA}

\newcommand{\UToledo}{\label{UToledo} University of Toledo, 2801 W. Bancroft St., Mail Stop 111, Toledo, OH, 43606}

\newcommand{\Toulouse}{\label{Toulouse} Universit\'{e} de Toulouse, UPS-OMP, IRAP, F-31028 Toulouse cedex 4, France}

\newcommand{\UBonn}{\label{UBonn} Argelander-Institut f\"ur Astronomie, Universit\"at Bonn, Auf dem H\"ugel 71, 53121 Bonn, Germany}

\newcommand{\UChile}{\label{UChile} Departamento de Astronom\'{i}a, Universidad de Chile, Camino del Observatorio 1515, Las Condes, Santiago, Chile}

\newcommand{\UConn}{\label{UConn} Department of Physics, University of Connecticut, Storrs, CT, 06269, USA}

\newcommand{\UCSD}{\label{UCSD} Center for Astrophysics and Space Sciences, Department of Physics,  University of California, San Diego, 9500 Gilman Drive, La Jolla, CA 92093, USA}

\newcommand{\UCSDAA}{\label{UCSDAA} Department of Astronomy \& Astrophysics,  University of California, San Diego, 9500 Gilman Drive, La Jolla, CA 92093, USA}

\newcommand{\UGent}{\label{UGent} Sterrenkundig Observatorium, Universiteit Gent, Krijgslaan 281 S9, B-9000 Gent, Belgium}

\newcommand{\ULyon}{\label{ULyon} Univ Lyon, Univ Lyon 1, ENS de Lyon, CNRS, Centre de Recherche Astrophysique de Lyon UMR5574, F-69230 Saint-Genis-Laval, France}

\newcommand{\UMass}{\label{UMass} University of Massachusetts—Amherst, 710 N. Pleasant Street, Amherst, MA 01003, USA}

\newcommand{\UWyoming}{\label{UWyoming} Department of Physics and Astronomy, University of Wyoming, Laramie, WY 82071, USA}

\newcommand{\LAM}{\label{LAM} Aix Marseille Univ, CNRS, CNES, LAM (Laboratoire d’Astrophysique de Marseille), Marseille, France}

\newcommand{\UHawaii}{\label{UHawaii} Institute for Astronomy, University of Hawaii, 2680 Woodlawn Drive, Honolulu, HI 96822, USA}

\newcommand{\UCM}{\label{UCM} Departamento de F\'{\i}sica de la Tierra y Astrof\'{\i}sica, Universidad Complutense de Madrid, E-28040, Spain}

\newcommand{\IPARC}{\label{IPARC} Instituto de F\'{\i}sica de Part\'{\i}culas y del Cosmos IPARCOS, Facultad de Ciencias F\'{\i}sicas, Universidad Complutense de Madrid, E-28040, Spain}
\newcommand{\UCT}{\label{UCT} Department of Astronomy, University of Cape Town, Rondebosch 7701, South Africa}

\newcommand{\STScI}{\label{STScI} Space Telescope Science Institute, 3700 San Martin Drive, Baltimore, MD 21218, USA}

\newcommand{\esaSTScI}{\label{esaSTScI} AURA for the European Space Agency (ESA), ESA Office, Space Telescope Science Institute, 3700 San Martin Drive, Baltimore, MD 21218, USA}

\newcommand{\McMaster}{\label{McMaster} Department of Physics and Astronomy, McMaster University, 1280 Main Street West, Hamilton, ON L8S 4M1, Canada}

\newcommand{\INAF}{\label{INAF} INAF -- Osservatorio Astrofisico di Arcetri, Largo E. Fermi 5, I-50157, Firenze, Italy}

\newcommand{\Sydney}{\label{Sydney} Sydney Institute for Astronomy, School of Physics A28, The University of Sydney, NSW 2006, Australia}

\newcommand{\UA}{\label{UA} Centro de Astronomía (CITEVA), Universidad de Antofagasta, Avenida Angamos 601, Antofagasta, Chile}

\newcommand{\CITA}{\label{CITA} Canadian Institute for Theoretical Astrophysics (CITA), University of Toronto, 60 St George St, Toronto, ON M5S 3H8, Canada}

\newcommand{\ASIAA}{\label{ASIAA} Institute of Astronomy and Astrophysics, Academia Sinica, No. 1, Sec. 4, Roosevelt Road, Taipei 10617, Taiwan}

\newcommand{\TKU}{\label{TKU} Department of Physics, Tamkang University, No.151, Yingzhuan Rd., Tamsui Dist., New Taipei City 251301, Taiwan}

\newcommand{\PSMA}{\label{PSMA} Penn State Mont Alto, 1 Campus Drive, Mont Alto, PA  17237, USA}

\newcommand{\ILL}{\label{ILL} Institut Laue-Langevin, 71 avenue des Martyrs, F-38042 Grenoble, France}

\newcommand{\TUM}{\label{TUM} Technical University of Munich, School of Engineering and Design, Department of Aerospace and Geodesy, Chair of Remote Sensing Technology, Arcisstr. 21, 80333 Munich, Germany}

\newcommand{\Surrey}{\label{Surrey} Department of Physics, University of Surrey, Guildford GU2 7XH, UK}

\newcommand{\Oxford}{\label{Oxford} Sub-department of Astrophysics, Department of Physics, University of Oxford, Keble Road, Oxford OX1 3RH, UK}

\newcommand{\AIP}{\label{AIP} Leibniz-Institut for Astrophysik Potsdam (AIP), An der Sternwarte 16, 14482 Potsdam, Germany}

\newcommand{\StAndrews}{\label{StAndrews} School of Physics and Astronomy, University of St Andrews, North Haugh, St Andrews, KY16 9SS, UK}

\newcommand{\IAC}{\label{IAC}{Instituto de Astrof\'isica de Canarias, C/ V\'ia L\'actea s/n, E-38205, La Laguna, Spain}}

\newcommand{\ULL}{\label{ULL}{Departamento de Astrof\'isica, Universidad de La Laguna, Av. del Astrof\'isico Francisco S\'anchez s/n, E-38206, La Laguna, Spain}}

\newcommand{\insubria}{ \label{insubria} Universit{\`a} dell’Insubria, via Valleggio 11, 22100 Como, Italy}

\newcommand{\Radcliffe}{\label{Radcliffe} Elizabeth S. and Richard M. Cashin Fellow at the Radcliffe Institute for Advanced Studies at Harvard University, 10 Garden Street, Cambridge, MA 02138, USA}

\newcommand{\UniCA}{\label{UniCA} Université Côte d'Azur, Observatoire de la Côte d'Azur, CNRS, Laboratoire Lagrange, 06000, Nice, France}

\newcommand{\Shizuoka}{\label{Shizuoka} Faculty of Global Interdisciplinary Science and Innovation, Shizuoka University, 836 Ohya, Suruga-ku, Shizuoka 422-8529, Japan}

\newcommand{\PMO}{\label{PMO} 
Purple Mountain Observatory, No. 10 Yuanhua Road, Qixia District, 210023 Nanjing, China}


\author{%
Miguel~Querejeta\inst{\ref{OAN}}              
\and Sharon~E.~Meidt\inst{\ref{UGent}}        
%
%
%
\and Yixian~Cao\inst{\ref{MPE}}
\and Dario~Colombo\inst{\ref{UBonn}} 
\and Eric~Emsellem\inst{\ref{ESO},\ref{ULyon}} %
\and Santiago Garc\'ia-Burillo\inst{\ref{OAN}} %
\and Ralf~S.~Klessen\inst{\ref{ITA},\ref{IWR},\ref{CfA},\ref{Radcliffe}}
\and Eric~W.~Koch\inst{\ref{NRAOAb}}
\and Adam~K.~Leroy\inst{\ref{OSU}}            
\and Marina Ruiz-Garc\'ia\inst{\ref{OAN},\ref{UCM}} %
\and Eva~Schinnerer\inst{\ref{MPIA}}          %
\and Rowan~Smith\inst{\ref{StAndrews}} 
\and Sophia Stuber\inst{\ref{MPIA}} %
\and Mallory Thorp\inst{\ref{UBonn}}
\and Thomas~G.~Williams\inst{\ref{Oxford}}
%
\and Médéric Boquien\inst{\ref{UniCA}}
\and Daniel~A.~Dale\inst{\ref{UWyoming}} 
\and Chris Faesi\inst{\ref{UConn}} 
\and Damian R. Gleis\inst{\ref{MPIA}} 
\and Kathryn~Grasha\inst{\ref{ANU}} 
\and Annie~Hughes\inst{\ref{IRAP}} 
\and Mar\'ia J. Jim\'enez-Donaire\inst{\ref{esaSTScI},\ref{OAN}}
\and Kathryn Kreckel\inst{\ref{Heidelberg}}
\and Daizhong~Liu\inst{\ref{PMO}} 
\and Justus Neumann\inst{\ref{MPIA}} %
\and Hsi-An~Pan\inst{\ref{TKU}} 
\and Francesca Pinna\inst{\ref{IAC},\ref{ULL}}
\and Alessandro Razza\inst{\ref{UChile}} 
\and Toshiki~Saito\inst{\ref{Shizuoka}}  %
\and Jiayi~Sun\inst{\ref{Princeton}}  
\and Antonio Usero\inst{\ref{OAN}} %
}

\institute{\OAN{} \and \UGent{} \and \MPE{} \and \UBonn{} \and \ESO{} \and \ULyon{} \and \ITA{} \and \IWR{} \and \CfA{} \and \Radcliffe{} \and \NRAOAb{} \and \OSU{} \and \UCM{} \and \MPIA{} \and \StAndrews{} \and \Oxford{} \and \UniCA{} \and \UWyoming{} \and \UConn{} \and \ANU{} \and \IRAP{} \and \esaSTScI{} \and \Heidelberg{} \and \PMO{} \and \TKU{} \and \IAC{} \and \ULL{} \and \UChile{} \and \Shizuoka{} \and \Princeton{}}

\date{Received ..... / Accepted .....}

\abstract {Spiral arms play a central role in disc galaxies, but their dynamical nature remains a long-standing open question. Azimuthal offsets between molecular gas and star formation are expected if gas crosses spiral arms, as predicted by quasi-stationary density wave theory. In this work, we measure offsets between CO and H$\alpha$ peaks in radial bins for 24 galaxies from the PHANGS survey that display a well-delineated spiral structure. The offsets exhibit substantial scatter, implying that star formation is not exclusively initiated at a coherent spiral shock. We define offsets such that positive values mean H$\alpha$ peaks lie ahead of CO peaks in the direction of galactic rotation. With this convention, 14 galaxies show mean positive CO-H$\alpha$ offsets, typically of a few hundred parsecs. In four of these 14 galaxies (17\% of the total), offsets become smaller with increasing radius, as expected for a single quasi-stationary spiral density wave. Ten galaxies (42\%) show positive mean offsets but no clear correlation with radius, which is compatible with multiple overlapping modes. In the remaining ten galaxies (42\%), we find no significantly positive offsets, which could point to transient dynamical spirals or material arms, where gas and stars co-rotate with the spiral perturbation. Across the full sample, we find mostly positive offsets between CO peaks and the gravitational potential minimum, confirming that gas often crosses the spiral perturbation. For the four galaxies with clear positive offsets and a radial trend, we derived pattern speeds in good agreement with the literature. Overall, our results suggest that even well-delineated spirals in the local Universe can arise from a variety of underlying dynamical mechanisms.
}


\titlerunning{Azimuthal offsets in spiral arms of nearby galaxies}
\authorrunning{M.~Querejeta et al.}

\maketitle 
\section{Introduction} 
\label{Sec:introduction}

Despite the widespread presence of spiral arms in the local Universe \citep[e.g.][]{2010ApJS..186..427N,2013MNRAS.435.2835W,2015ApJS..217...32B}, understanding their origin and nature remains an open subject of research in contemporary astrophysics. One of the most popular theories to explain the spiral pattern in disc galaxies is quasi-stationary spiral density wave theory \citep{1964ApJ...140..646L,1966PNAS...55..229L,1989ApJ...338...78B,2000dyga.book.....B},
which posits that spiral arms are long-lived, self-sustaining modes of the disc that rigidly rotate at a constant pattern speed, $\Omega_{\rm P}$. In this model, spiral arms are local overdensities formed by different stars over time as gas and stars cross the spirals.

In another leading theory, spirals are temporary material patterns that rotate with the disc. This would be the case for spirals originating through swing amplification 
\citep{1965MNRAS.130..125G,1966ApJ...146..810J,2024ApJ...966...62M} or spirals that are tidal features produced during interactions and mergers \citep[e.g.][]{1969ApJ...158..899T,1981seng.proc..111T,1994A&A...290..785D,2014MNRAS.443.2757K}. The former are thought to be consistent with the spirals in flocculent galaxies, whereas the latter are argued to be prominent in some grand-design spirals, such as M51 (especially in the outer regions; \citealt{2009ApJ...702..277M}, \citealt{2020MNRAS.492.2973T}). In either case, in contrast to the Lin-Shu quasi-stationary modes, material spirals are temporary features of the disc, and they co-rotate with the disc material \citep{2014PASA...31...35D,2022ARA&A..60...73S} such that the gas and stars do not cross the spiral.

The spirals formed in modern simulations of isolated disc galaxies exhibit properties resembling both types of theory: They are short-lived rigidly rotating `groove' modes of the disc shaped by both swing amplification (which amplifies small perturbations as they shear due to differential rotation) and changes to the disc properties brought about by earlier spiral features (\citealt{2014ApJ...785..137S}; see also \citealt{2020MNRAS.496..767B,2019ApJ...884....3S,2024ApJ...966...62M,2024MNRAS.528.5286H}).  These spirals are often composed of multiple independent modes, each excited around its own co-rotation. Observational measurements of multiple independent pattern speeds \citep[e.g.][]{2008ApJ...688..224M,2009ApJ...702..277M,2011ApJ...741L..14F,2014ApJS..210....2F,2024ApJ...966..110F,2014ApJ...784....4C,2024A&A...686L..14M} tend to support this picture, especially when the co-rotation resonance of one spiral overlaps with a resonance of another spiral \citep{2014ApJ...785..137S}. Resonance overlaps such as these, which are also thought to allow for non-linear mode coupling \citep{1988MNRAS.232..733S}, produce radially extended spiral arms that allow for energy and angular momentum transfer extending widely across the disc \citep{1997A&A...322..442M}.

The spirals produced in these different pictures yield different responses to the gas and the young and old stellar discs, which can offer a way to discriminate between them. In the picture of quasi-stationary density waves, the shock predicted in the gas as it overtakes the spiral pattern leads to a systematic offset between the accumulation of gas at the shock front and star formation peaks and thus also between young and older stars \citep{1969ApJ...158..123R}. This offset will decrease in amplitude moving outwards along the spiral and switch direction, passing from inside to outside co-rotation. For multiple `groove' modes that extend only out to co-rotation, no switch in offset location will be produced. For material arms, no systematic offset is expected \citep{2012MNRAS.426..167G,2012MNRAS.421.1529G,2014MNRAS.443.2757K,2015PASJ...67L...4B}.

Numerous studies have attempted to identify angular offsets associated with spiral arms. Different approaches include measuring changes in pitch angle for several bands or colour maps \citep{2009A&A...499L..21G,2020MNRAS.496.1610A,2023MNRAS.524...18M,2024MNRAS.527L..66M},
identifying stellar populations or clusters of different ages \citep{2015ApJ...810....9C,2017ApJ...845...78C,2018MNRAS.478.3590S,2019NatAs...3..178P},
pinpointing offsets involving HII regions \citep{2009ApJ...697.1870E,2017MNRAS.465..460E,2013A&A...560A..59C},
deriving w cross-correlations for different tracers \citep{2008AJ....136.2872T,2011ApJ...735..101F,2013ApJ...779...42S},
and even conducting comparisons with numerical simulations \citep{2007MNRAS.376.1747D,2015ApJ...800..106W}.

Based on the analysis of offsets or pitch angles, some studies have found strong evidence supporting quasi-stationary spiral density wave theory for certain galaxies \citep[e.g.][]{2009ApJ...697.1870E,2019NatAs...3..178P,2019ApJ...874..177M,2020MNRAS.496.1610A,2023MNRAS.524...18M,2024MNRAS.527L..66M}.
However, such offsets have not been found in other galaxies \citep[e.g.][]{2009ApJ...697.1870E,2011ApJ...735..101F,2012MNRAS.424.1636F,2013ApJ...765..105M,2015ApJ...810....9C,2024MNRAS.534..883C,2024MNRAS.528.1276M}.
Thus, the picture that emerges so far is that quasi-stationary density wave theory may apply to some spiral galaxies but not to others. The choice of tracers and methodology is critical, as \citet{2013ApJ...763...94L} have shown that apparent discrepancies in the nature of gas-star formation offsets in M51 \citep{2008AJ....136.2872T,2009ApJ...697.1870E,2011ApJ...735..101F} are due to the choice of tracer, specifically because HI enhancements in M51 do not trace compressed gas as much as a product of H$_2$ photodissociation close to star-forming regions. The conclusion of \citet{2013ApJ...763...94L} was that CO and H$\alpha$ are the optimal tracers when it comes to establishing offsets between gas and star formation in spiral arms. Another important caveat is resolution, as small offsets will be washed out if observations do not attain sufficient physical resolution. In this sense, \citet{2009ApJ...697.1870E} detected offsets in some galaxies but not in others and hypothesised that the limited resolution (${\sim}500$\,pc) and/or sensitivity of their CO data could explain the lack of measured offsets in some cases. In other words, they have suggested that spiral density wave theory might apply across their sample, but, possibly, the signature only becomes apparent for the galaxies with sufficient resolution and sensitivity. This motivates the interest of revisiting this problem with high-resolution and high-sensitivity CO and H$\alpha$ data for a sample of nearby galaxies.

The main goal of this paper is to quantify CO-H$\alpha$ offsets in the PHANGS\footnote{Physics at High Angular resolution in Nearby GalaxieS; \url{http://www.phangs.org}} sample of nearby galaxies.
PHANGS provides the largest and highest resolution sample of nearby spiral galaxies with CO data to carry out this experiment  \citep{2021ApJS..257...43L}, with an approximately five times higher physical resolution than \citet{2009ApJ...697.1870E}.
The structure of this paper is the following. In Sect.~\ref{Sec:method} we present the methodology and outline the main expectations for different scenarios that explain spiral structure. In Sect.~\ref{Sec:data} we describe our tracers of molecular gas and star formation. The main results of the paper are presented in Sect.~\ref{Sec:results} and discussed in Sect.~\ref{Sec:discussion}. We finish with a summary in Sect.~\ref{Sec:concl}.

\begin{figure}[!ht]
\begin{center}
\includegraphics[trim=0 0 0 0, clip,width=0.45\textwidth]{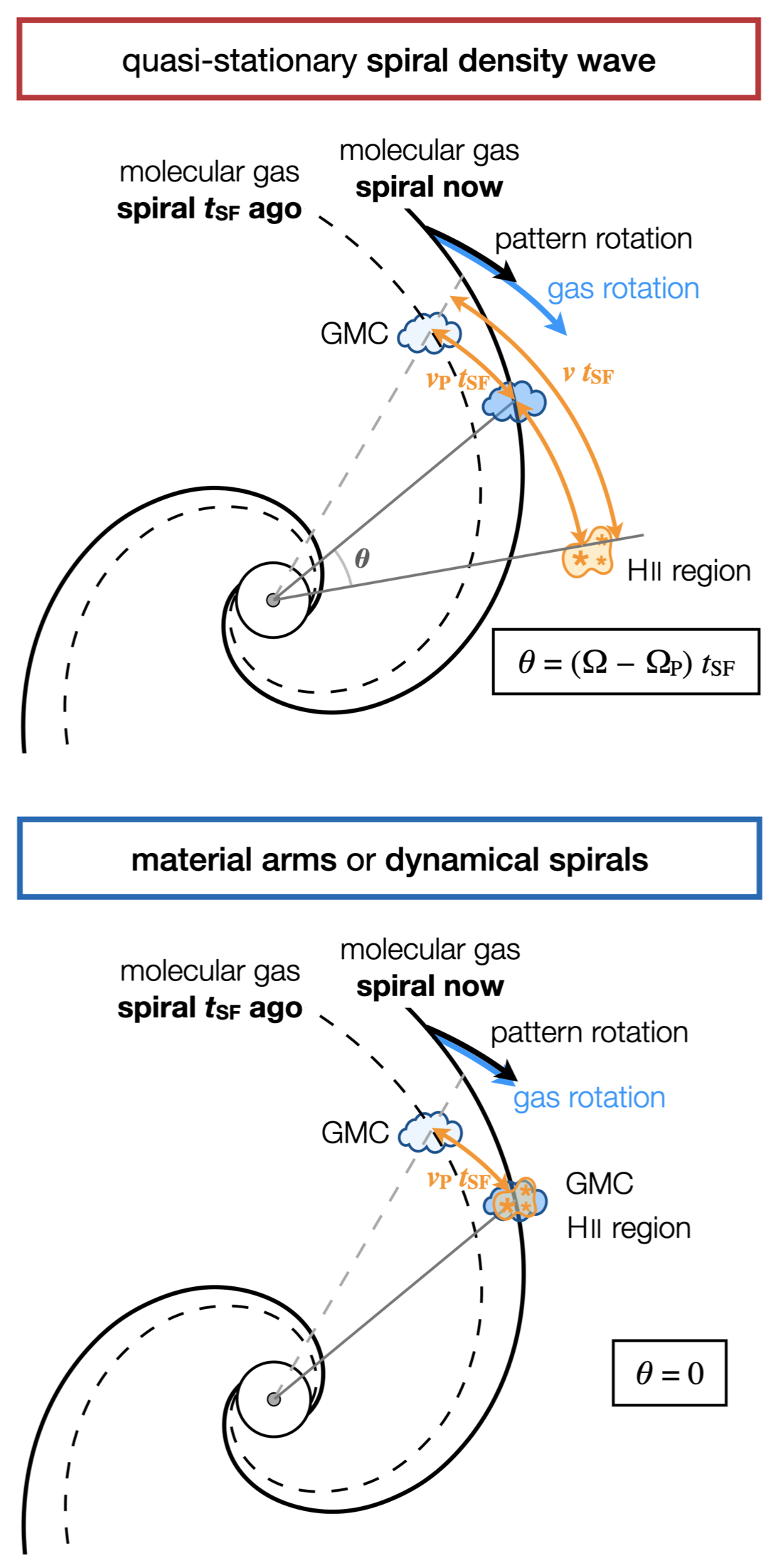}
\end{center}
\caption{\mod{Top panel: Quasi-stationary spiral density wave theory predicts azimuthal offsets in spiral arms}. The thick solid black line shows the current position of the molecular gas spiral, while the dashed line displays its position, $t_{\rm SF}$, earlier. If a shock coherently triggers the collapse of molecular gas along the spiral and if it takes $t_\mathrm{SF}$ for the resulting HII regions to be visible, the peaks of star formation \mod{will be displaced `downstream' inside co-rotation. Measuring offsets at different galactocentric radii, one can calculate the (constant) pattern speed of the spiral and star formation timescale $t_\mathrm{SF}$ \citep{2009ApJ...697.1870E}.} \mod{If multiple overlapping spiral modes are present in the disc, the situation illustrated in the top panel will apply for a limited radial range, with different pattern speeds operating at different radial zones. Bottom panel: Situation expected for material arms or dynamical spirals, where the spiral rotates at the same angular speed as the underlying disc material and therefore no bulk offsets are expected.} In real galaxies, the fact that star formation is not limited to a coherent spiral shock introduces significant scatter in the offset measurements \mod{in both cases}.
}
\label{fig:spiral_sketch}
\end{figure}

\section{Methodology and expectations} 
\label{Sec:method}

\begin{figure*}[!ht]
\begin{center}
\includegraphics[trim=0 0 0 0, clip,width=0.95\textwidth]{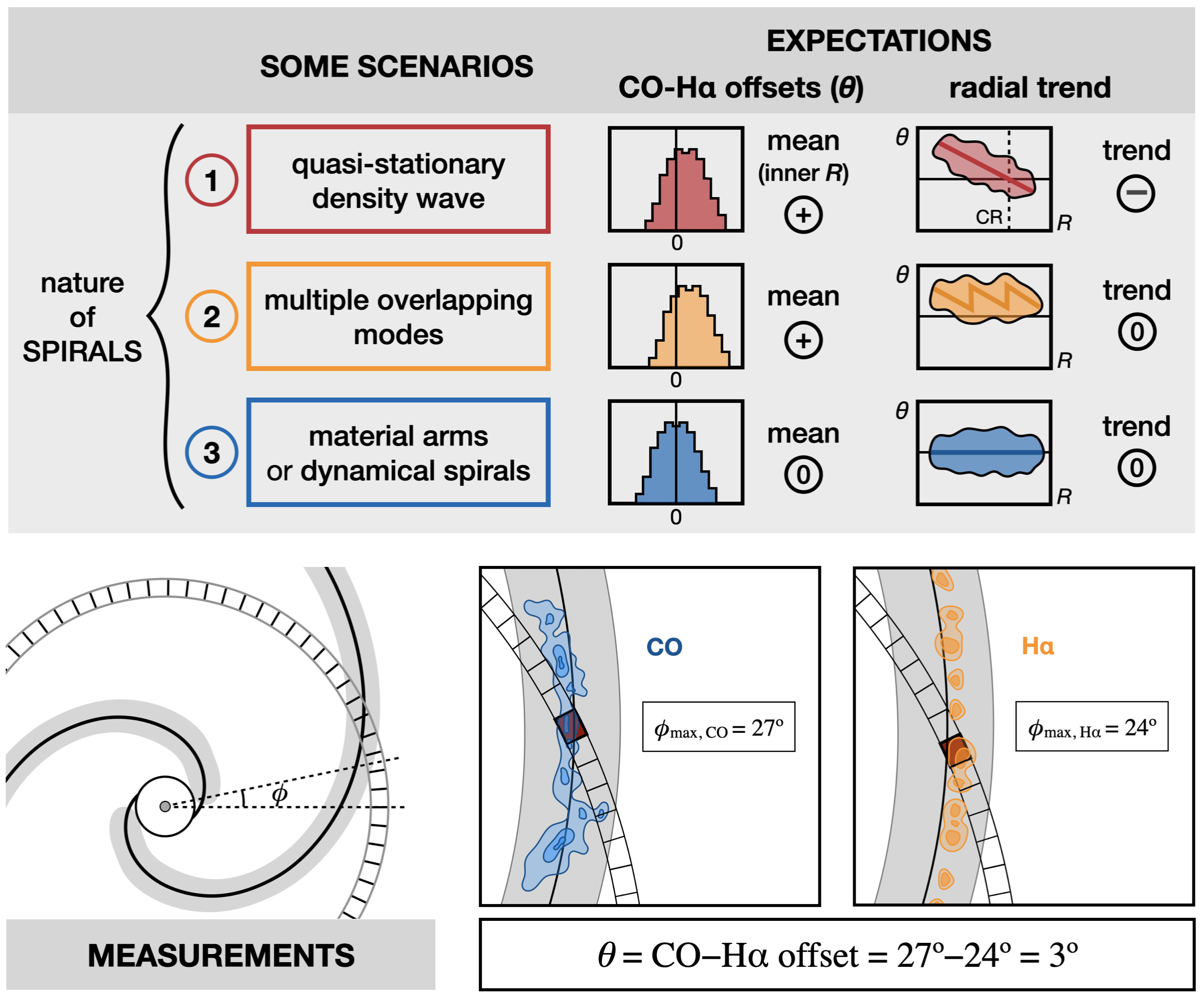}
\end{center}
\caption{Top: Illustration showing the expectations for a non-exhaustive list of three scenarios that explain spiral structure: a quasi-stationary density wave (constant pattern speed), multiple overlapping modes (i.e.\ different pattern speeds across the disc), and material arms (always made up of the same stars, all of which are by definition co-rotating) or dynamical spirals (transient features made up of rather short, unconnected segments). Middle panels: Expected histograms for the distribution of CO-H$\alpha$ offsets ($\theta$) as well as the expected mean offset. We define offsets such that positive values mean H$\alpha$ peaks lie ahead of CO peaks in the direction of galactic rotation. For quasi-stationary density waves, we expect such offsets to be positive inside co-rotation and negative outside; however, since we sample mostly the inner parts of galaxies, the net offsets should typically be positive within our fields of view. The right panels indicate whether we should find a significant correlation between offsets and radius or not. Bottom: Illustration of how we measure azimuthal offsets in spiral arms in practice. For each radial bin (an elliptical annulus of width ${\sim}100$\,pc that accounts for the inclination and PA of the galaxy), we average the CO and H$\alpha$ intensities in azimuthal bins. We then find the azimuth ($\phi$) corresponding to the maximum of both tracers within the footprint of each spiral mask ($\phi_{\rm max}$). The angular difference between the bins of peak intensity yields the offset of interest, $\theta$, at that radius. To account for galaxy rotation, if the galaxy rotates anti-clockwise, the offset $\theta$ is multiplied by $\eta = -1$.
}
\label{fig:method}
\end{figure*}

According to the seminal work by \citet{1969ApJ...158..123R}, spiral arms should result in a shock that triggers star formation as gas clouds get compressed \citep[see also][]{1975ApJ...196..381R,2004MNRAS.349..909G}. In the idealised case where star formation is exclusively initiated at the spiral shock, under spiral density wave theory we would expect \mod{systematic} offsets between gas and star formation (except at co-rotation), as material travels a certain distance from the collapse of gas until star formation tracers are visible (e.g.\ H$\alpha$). 

The ideal picture of star formation exclusively triggered at spiral arms is ruled out by observations, because stars also form in interarm regions \citep{2010ApJ...725..534F,2021A&A...656A.133Q,2024A&A...687A.293Q,2022ApJ...941L..27W} 
and star formation in spirals is not only initiated by a coherent shock, but also elongated structures such as spurs and feathers play an important role \citep{2017ApJ...836...62S}. Furthermore, turbulence and feedback effects also modify the state of molecular gas locally and affect its ability to collapse and form stars \citep[e.g.][]{2015MNRAS.450.4035F,2015arXiv151103457K,2025arXiv250303305C}.
Yet, stellar spiral arms do concentrate gas and, indirectly, star formation \citep{1988Natur.334..402V,1993A&A...274..148G,2020ApJ...892..148S,2021A&A...656A.133Q,2024A&A...687A.293Q,2024ApJ...973..137S}. This means that while the departures from the ideal picture will induce scatter on the measurements, we do expect positive average offsets between CO and H$\alpha$ if gas crosses spiral arms. This is because more gas is accumulated and, on average, more star formation is initiated in the spiral region.

According to this picture, following quasi-stationary density wave theory, in the inner parts of galaxies H$\alpha$ peaks should be statistically displaced with respect to CO peaks in the direction of rotation. If the spiral pattern (density wave) rotates as a rigid structure with a constant pattern speed, these offsets are expected to decrease with increasing radius, all the way out to co-rotation, where the relative ordering of the tracers should flip. Following \citet{2009ApJ...697.1870E}, Fig.~\ref{fig:spiral_sketch} illustrates how offsets at a certain galactocentric radius will depend on the disc rotation at that radius (angular speed of the disc material) compared to the rotation of the spiral perturbation (assumed to be a fixed pattern speed). 
Thus, measuring these offsets for different galactocentric radii allows for simultaneous fitting of the constant pattern speed, $\Omega_{\rm P}$, and the timescale of star formation, $t_{\rm SF}$, which is also assumed to be constant. This is possible when plotting the offsets as a function of the angular frequency, $\Omega = v_{\rm rot} / R$, which requires adopting a certain rotation curve, $v_{\rm rot} (R)$. Numerical simulations relying on a fixed spiral potential confirm the expectation that the azimuthal offsets between tracers of gas and star formation should vary with galactocentric radius \citep{2004MNRAS.349..909G,2015PASJ...67L...4B}, but the result for more flocculent models without a fixed and strong spiral potential is less clear \citep[e.g.][]{2011ApJ...735..101F}.

There are three basic scenarios for spiral structure that we would like to differentiate, and where the offset measurements can be informative, as illustrated by Fig.~\ref{fig:method}. The first one is the traditional Lin-Shu spiral density wave with a constant pattern speed across the disc; in this case, we expect to measure CO-H$\alpha$ offsets which vary radially (showing a linear trend with $\Omega$) across the whole disc. The second scenario involves multiple overlapping rigid spirals, each with their own pattern speed, which result in a positive mean CO-H$\alpha$ offset (since disc material travels across the spiral but it is always inside co-rotation); however, we would not expect a consistent trend with radius (or $\Omega$) across the disc. Finally, for material arms or dynamical spirals, offsets should randomly scatter around zero since the material does not cross the spiral pattern, resulting in a zero mean offset.

Figure~\ref{fig:method} also shows schematically how we measure azimuthal offsets in spiral arms. First, we define a set of bins in fixed intervals of galactocentric radius, which results in elliptical annuli in the plane of the sky (according to the disc inclination and position angle from the PHANGS sample table 1.6; \citealt{2020ApJ...897..122L}). The width of the radial bins is chosen to be the largest between the resolution of the CO and H$\alpha$ map; in most cases, this is limited by the beam of the CO map, which is typically ${\sim} 1\arcsec$ (${\sim}100$\,pc at a distance of ${\sim}20$\,Mpc). Within each radial bin, we define a number of boxes at azimuthal intervals matching the same physical resolution (the azimuthal angle is measured anti-clockwise in the plane of the galaxy). This means that closer to the centre, the number of boxes will be smaller, while at larger radii there will be many more boxes. Within each of these boxes, typically ${\sim}100\, \mathrm{pc} \times 100 \, \mathrm{pc}$, we calculate the average intensity on the CO and H$\alpha$ maps. Then, for each galactocentric elliptical annulus and within the footprint of each spiral mask, we identify the azimuth of the boxes with maximum mean intensity in CO and H$\alpha$, respectively. This way, we derive the angular offset between CO and H$\alpha$ ($\phi_{\rm max}^{\rm CO}-\phi_{\rm max}^{\rm H\alpha}$) in the plane of the galaxy for each spiral arm as a function of galactocentric radius. To account for the direction of rotation of the galaxy, we multiply the offsets by $\eta = +1$ if the galaxy rotates clockwise, or $\eta = -1$ if the galaxy rotates anti-clockwise (assuming spiral arms are trailing). This is because we adopt the definition that positive offsets imply that H$\alpha$ peaks lie ahead of CO peaks in the direction of galactic rotation. 

The expectation for offsets with respect to the potential minimum is less clear. According to \citet{1969ApJ...158..123R}, the shock is located immediately before the potential minimum. Thus, for the inner parts of galaxies, the order of tracers should be CO peaks, then stellar mass peaks, and finally H$\alpha$ peaks, as we move from the concave to the convex part of the spiral (assuming the spirals are trailing). Yet, even in simulations with a rigidly rotating potential, the relative offset between gas peaks and the gravitational potential minimum is not as straightforward and depends on the sound speed of the gas \citep{2007PhDT.........1D,2008ApJ...675..188W}. For completeness, we also report on offsets between the stellar mass peaks and CO peaks in Sect.~\ref{Sec:NIR-CO}.

\section{Sample and data} 
\label{Sec:data}

We focus on the 23 spiral galaxies from PHANGS--ALMA \citep{2021ApJS..257...43L} with a spiral mask in \citet{2021A&A...656A.133Q}; \mod{a CO map that extends beyond the end of the bar to cover, at least partially, the spiral arms (Sect.~\ref{Sec:PHANGS--ALMA}); and an H$\alpha$ image available from PHANGS} (Sect.~\ref{Sec:SFRs}). We added M51 (NGC\,5194), which was targeted by similar studies in the past \citep[e.g.][]{2017MNRAS.465..460E}. As explained in \citet{2021A&A...656A.133Q}, we emphasise that this sample is biased towards grand-design spiral galaxies, as it mostly excludes flocculent discs, where the spiral structure is less clear. We carried out radially binned measurements of the angular offsets as explained in Sect.~\ref{Sec:method}. We also employed stellar mass maps tracing the potential minimum (Sect.~\ref{Sec:stellarmass}). As a sanity check, we considered JWST maps at 2\,$\mu$m for \mod{some} galaxies as an alternative tracer of stellar mass (Sect.~\ref{Sec:JWST}). We measured the offsets within the footprint of the spiral masks presented in Sect.~\ref{Sec:masks}.

\subsection{CO as molecular gas tracer} 
\label{Sec:PHANGS--ALMA}

We rely on ${\sim} 1\arcsec$-resolution (${\sim}100$\,pc) $^{12}$\mbox{CO(2--1)} intensity maps from PHANGS--ALMA as a molecular gas tracer, which account for ${\sim} 70\%$ of all the CO emission in the galaxies \citep{2021ApJS..255...19L}. We consider the zeroth-order (integrated intensity) CO moment maps using the `broad' masks presented in \citet{2021ApJS..255...19L} and publicly released in July 2022 (PHANGS--ALMA version~4.0). This provides high CO flux completeness, which makes it ideal for our purpose of determining offsets between peaks. We did not convert the CO integrated intensities to actual surface densities, as we are only interested in azimuthal offsets at fixed galactocentric radii and assume that $\alpha_{\rm CO}$ or $R_{21}$ line ratio variations at fixed radius and within the spiral region are negligible.

As a point of comparison with previous studies, we also considered M51 (NGC\,5194) through the PAWS $^{12}$\mbox{CO(1--0)} dataset \citep{2013ApJ...779...42S,2013ApJ...779...43P}, using the publicly delivered zeroth-order moment map. PAWS provides a resolution of ${\sim} 1''$ (${\sim}40$\,pc at the distance of M51).

\subsection{H$\alpha$ as star formation tracer} 
\label{Sec:SFRs}

Razza et al.\ (in prep.) obtained ground-based narrow-band H$\alpha$ images for a total of 65 galaxies from the PHANGS--ALMA parent sample. These were observed with the Wide Field Imager (WFI) at the MPG \mbox{2.2-metre} telescope in La Silla or with the DirectCCD camera at the du~Pont \mbox{2.5-metre} telescope in Las Campanas between 2016 and 2019 (these maps have been used e.g.\ in \citealt{2022ApJ...927....9P}). Both broad-band and narrow-band images were obtained, which allowed us to derive H$\alpha$ continuum-subtracted images. The resolution, limited by seeing, is typically ${\sim} 1\arcsec$ (full width at half maximum, FWHM), ranging from $0.6\arcsec$ to $1.3\arcsec$.

Following \citet{2009ApJ...697.1870E}, by default we consider offsets associated with H$\alpha$ flux peaks along spirals (which corresponds to a star-formation timescale of ${\sim}5{-}10$\,Myr; \citealt{2022MNRAS.516.3006K}). We did not apply any specific recipes to transform H$\alpha$ fluxes to actual star formation rates. While we are not concerned about the actual value of SFR, we are interested in locating the azimuthal peak of star formation around spiral arms, \mod{which} could in principle be affected by extinction (we discuss this caveat in Sect.~\ref{Sec:caveats}). 

\subsection{{\it Spitzer} IRAC $3.6$\,$\mu$m to trace stellar mass} 
\label{Sec:stellarmass}

In order to identify the minima of the gravitational potential, we used a set of stellar mass maps. For most galaxies, these are based on {\it Spitzer} $3.6$\,$\mu$m imaging obtained from the {\it Spitzer} Survey of Stellar Structure in Galaxies (S$^4$G; \citealt{2010PASP..122.1397S}), corrected for dust emission using an independent component analysis (ICA) technique (similar to principal component analysis, PCA). The method, introduced in \citet{2012ApJ...744...17M}, was applied to the entire S$^4$G sample in \citet{2015ApJS..219....5Q}. It makes use of simultaneous $3.6$\,$\mu$m and 4.5\,$\mu$m observations and exploits the fact that stellar and dust emission are expected to have very different [3.6]-[4.5] colours. Then, ICA identifies the global stellar and dust [3.6]-[4.5] colours that best describe the two underlying components. For the galaxies which are not in S$^4$G, we used the original IRAC $3.6$\,$\mu$m maps (not corrected with ICA; see \citealt{2021A&A...656A.133Q} for details). We do not care about the actual conversion of NIR fluxes into stellar mass, since we only need to identify the local peaks, so we do not discuss the details of the mass-to-light (M/L) ratio. However, we note that a constant M/L on the ICA-corrected $3.6$\,$\mu$m maps is expected to yield a good representation of the stellar surface density according to \citet{2014ApJ...788..144M}.

\subsection{Alternative tracers based on JWST imaging} 
\label{Sec:JWST}

\mod{In spiral arms, the stellar mass maps from Sect.~\ref{Sec:stellarmass} rely strongly on the ICA correction for dust emission, since the additional non-stellar emission is especially significant around star-forming regions.
For this reason, as a sanity check, for a few galaxies} we considered JWST maps at 2\,$\mu$m as an alternative tracer of stellar mass. PHANGS-JWST observations are available for multiple bands, including NIRCam F200W, as part of a Cycle 1 JWST Treasury program (proposal 2107; PI: J.\ Lee). \mod{From our sample of spiral galaxies, those Cycle 1 observations cover NGC\,0628, NGC\,1300, NGC\,1385, NGC\,1512, NGC\,1566, NGC\,1672, NGC\,3627, NGC\,4303, NGC\,4321, NGC\,4535.} (See \citet{2023ApJ...944L..17L} and \citet{2024ApJS..273...13W} for additional information about the JWST observations and data reduction.)

\subsection{Spiral masks}
\label{Sec:masks}

Throughout this paper, we use the spiral component of the environmental masks defined in \citet{2021A&A...656A.133Q}. These masks delimit morphological features visually identified on NIR images which, in addition to spiral arms, include centres, bars, rings, lenses, and discs. For M51 (NGC\,5194), we constructed a \mod{new} spiral mask in full analogy to the masks released for PHANGS \citep{2021A&A...656A.133Q}.

Specifically, spiral masks were constructed following three steps. Firstly, regions of bright $3.6$\,$\mu$m emission were identified along each spiral arm (using an unsharp-mask approach) and fitted with an analytic log-spiral function in the plane of the galaxy. Secondly, analytic log-spiral curves were assigned a width determined empirically based on CO emission. Finally, the resulting masks were visually inspected and the starting and ending point of some segments were extended in order to enforce continuity. For most galaxies, the NIR images and the analytic log-spiral fits come from S$^4$G \citep{2015A&A...582A..86H}, but we also relied on archival {\it Spitzer} $3.6$\,$\mu$m observations in some cases, where we also performed the log-spiral fitting, as explained in \citet{2021A&A...656A.133Q}.

The spiral masks from \citet{2021A&A...656A.133Q} consist of smooth, dilated log-spiral segments, with a typical width of ${\sim}1{-}2$\,kpc in order to accommodate most $3.6$\,$\mu$m, CO, and H$\alpha$ emission along the arm. Locally, the distribution of molecular gas or star formation in the arms looks often thinner, but the presence of kinks and irregularities requires this width in order to warrant full coverage along the entire arm.

\section{Results} 
\label{Sec:results}

We start by presenting our measurements of the CO--H$\alpha$ offsets in Sect.~\ref{Sec:CO-Ha}. In Sect.~\ref{Sec:radial_trend} we examine any radial trends present in the offsets. Then, we consider the offsets with respect to the peak in stellar mass in Section~\ref{Sec:NIR-CO}. In Sect.~\ref{Sec:OmegaP} we take one step further and, for the galaxies with offsets that follow a consistent radial trend, we derive the star formation timescales and pattern speeds following \citet{2009ApJ...697.1870E}.

\subsection{Offsets between molecular gas and star formation} 
\label{Sec:CO-Ha}

\begin{figure*}[!ht]
\begin{center}
\includegraphics[trim=0 0 0 0, clip,width=0.95\textwidth]{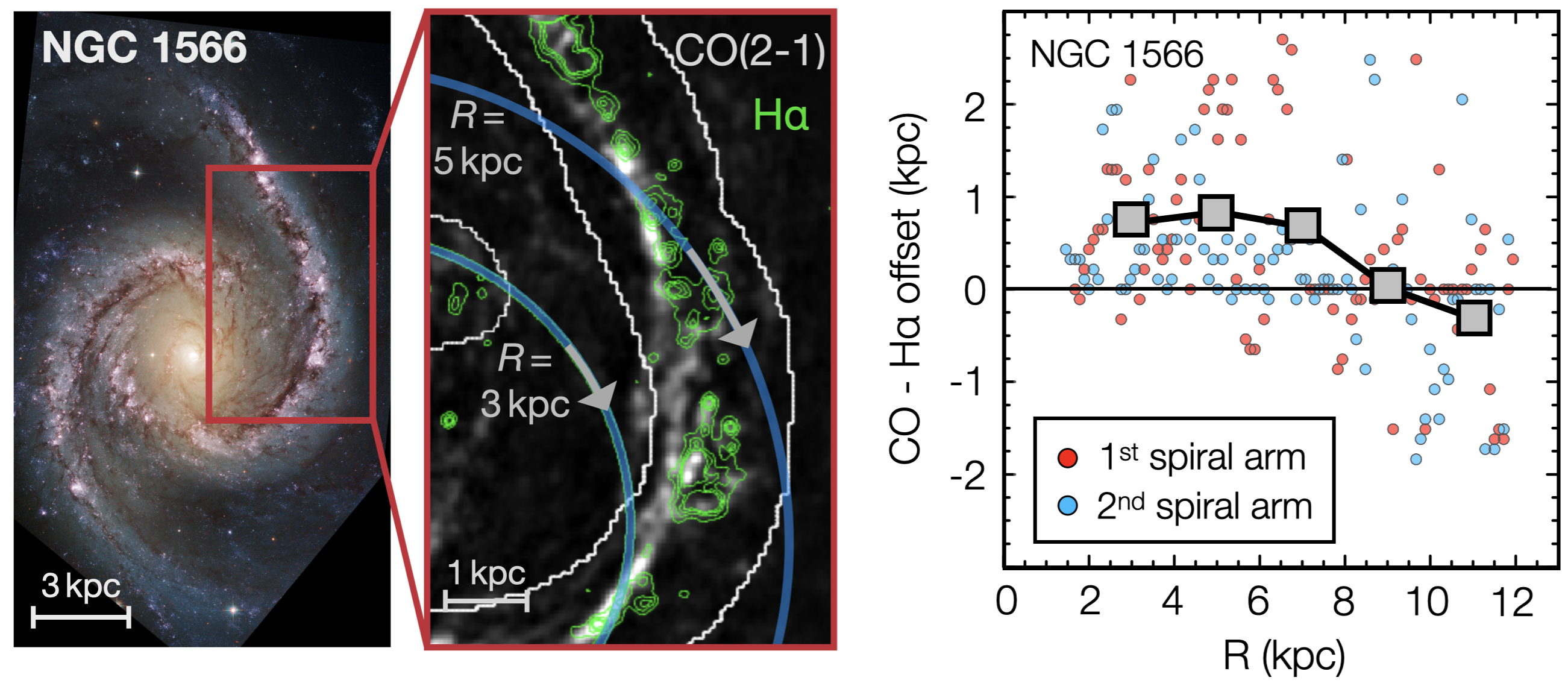}
\end{center}
\caption{Colour composite image of NGC\,1566 based on six different HST WFC3/UVIS filters \citep{2023ApJ...944L..17L}. The middle panel illustrates offsets in the western arm of NGC\,1566 between molecular gas (CO, greyscale on background image) and star formation (H$\alpha$, green contours). Blue shaded ellipses indicate two reference radial bins ($R=3$\,kpc and $R=5$\,kpc), the arrows indicate the sense of rotation, and white contours delimit the spiral mask. 
In the right panel, the radial distribution of all CO-H$\alpha$ offsets across the entire NGC\,1566 are shown. The first spiral arm is the eastern arm; the second spiral arm is the western arm. \mod{The error bars are comparable to the size of the plotted circles.} The gray squares show running means at radial intervals of 2\,kpc (combining both arms).
}
\label{fig:NGC1566zoom}
\end{figure*}

Figure~\ref{fig:NGC1566zoom} illustrates the CO-H$\alpha$ offsets in NGC\,1566. As expected, if gas and stars cross the spiral perturbation, star formation is displaced towards the convex (leading) side of the western spiral compared to molecular gas. Yet, this displacement is far from constant. At some radial bins, HII regions are nearly cospatial with a CO peak (offset $\sim 0$), while most bins show typical offsets of a few hundred parsecs, up to a couple of kiloparsecs in the most extreme cases (e.g.\ at $R \sim 5$\,kpc). By default we measure offsets between CO and H$\alpha$ peaks, but in Appendix~\ref{sec:appendix_robustness} we consider instead mean intensity-weighted positions, which decreases the scatter, but conclusions are not affected. The right panel demonstrates how, despite the significant scatter, a bulk positive offset emerges for the inner radii, and the mean offset traced by the running means declines until a radius of $R \sim 9$\,kpc, after which it becomes slightly negative. We opt for running means instead of running medians because we are interested in the bulk average offset with respect to zero, and not the typical offset at a given radius.

\begin{figure*}[t]
\sidecaption
\includegraphics[trim=0 0 0 0, clip,width=11.5cm]{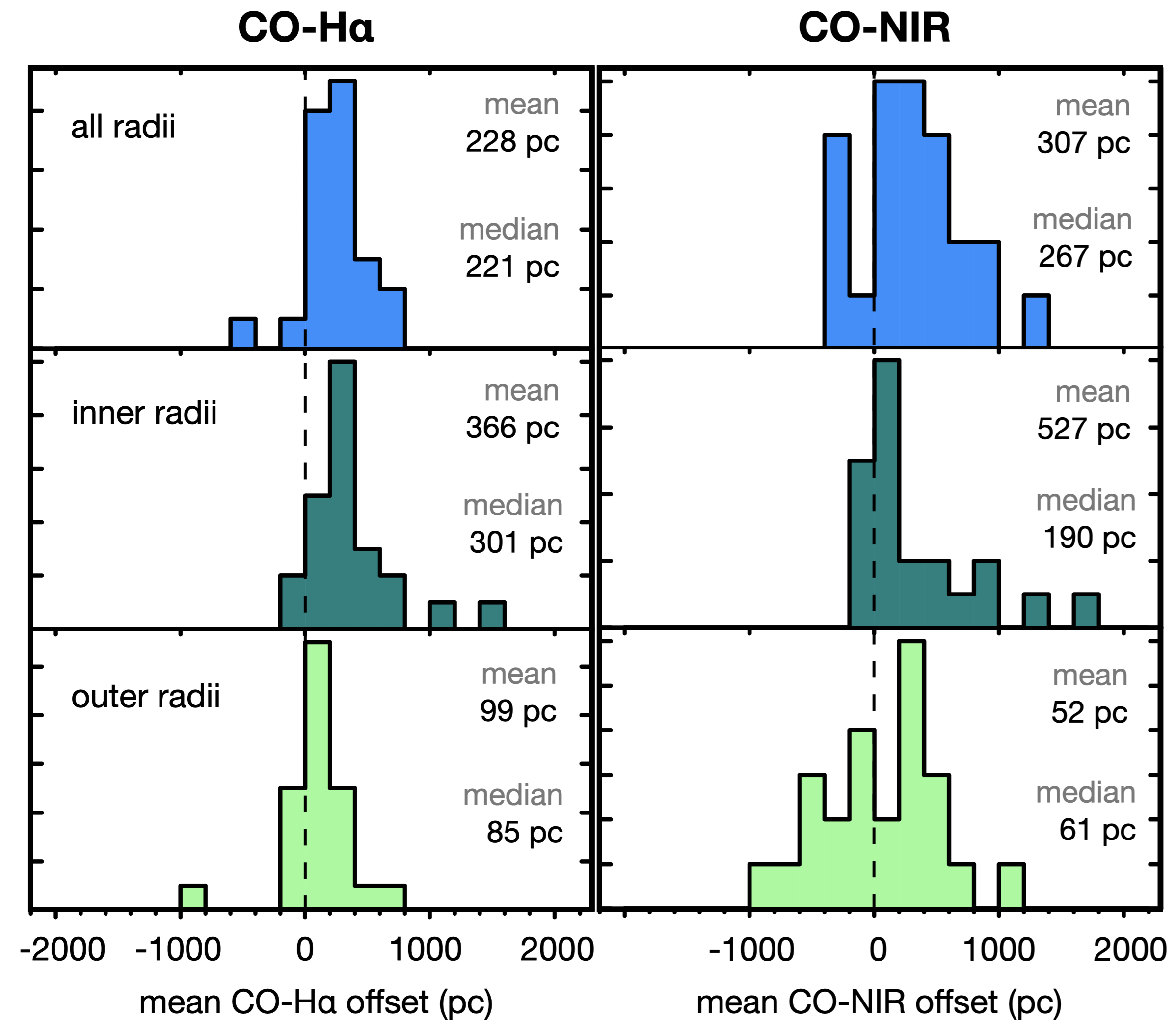}
\caption{Histograms showing the distribution of the mean CO-H$\alpha$ and CO-NIR offsets across the whole galaxy sample. Each histogram includes one mean offset per galaxy, computed by averaging all measured offsets for all radii (top), the inner half (middle, $R < (R_{\rm min}^{\rm CO} + R_{\rm max}^{\rm CO})/2$), and the outer half (bottom, $R \ge (R_{\rm min}^{\rm CO} + R_{\rm max}^{\rm CO})/2$)). }
\label{fig:histo_CO-Ha}
\end{figure*}

Most galaxies in our sample show positive mean CO-H$\alpha$ offsets in the plane of the galaxy, as illustrated in Fig.~\ref{fig:histo_CO-Ha}, which displays the statistical results for the whole sample. 
For each galaxy we calculate the average of all CO-H$\alpha$ offsets, considering all spiral arms simultaneously; the numerical results per galaxy are listed in Table~\ref{table:stats}.
The typical mean offsets per galaxy are of several hundred parsecs (mean $230$\,pc, median $220$\,pc), ranging from $-400$\,pc up to $740$\,pc. As a first simple test on whether offsets decrease with galactocentric radius, in Fig.~\ref{fig:histo_CO-Ha} and Table~\ref{table:stats} we also split into the innermost and outermost half radii for each galaxy ($R < (R_{\rm min}^{\rm CO} + R_{\rm max}^{\rm CO})/2$, and $R \ge (R_{\rm min}^{\rm CO} + R_{\rm max}^{\rm CO})/2$, respectively). While an arbitrary cut, this division maximises the number of datapoints in two radial bins, which allowed us to check if a radial trend is present despite large scatter. Overall, the mean offsets become closer to zero or even negative \mod{from the inner to the outer} radii (from an average of $370$\,pc down to $100$\,pc). In Sect.~\ref{Sec:radial_trend} we examine the radial dependence of offsets in more detail.
For completeness, the right panels of Fig.~\ref{fig:histo_CO-Ha} show the offsets between molecular gas and stellar mass peaks (CO-NIR), which follow a similar behaviour as the CO-H$\alpha$ offsets albeit with more scatter; we focus on this in Sect.~\ref{Sec:NIR-CO}.

\subsection{Radial variation in the offsets} 
\label{Sec:radial_trend}

\begin{figure*}[!ht]
\begin{center}
\includegraphics[trim=0 0 0 0, clip,width=0.95\textwidth]{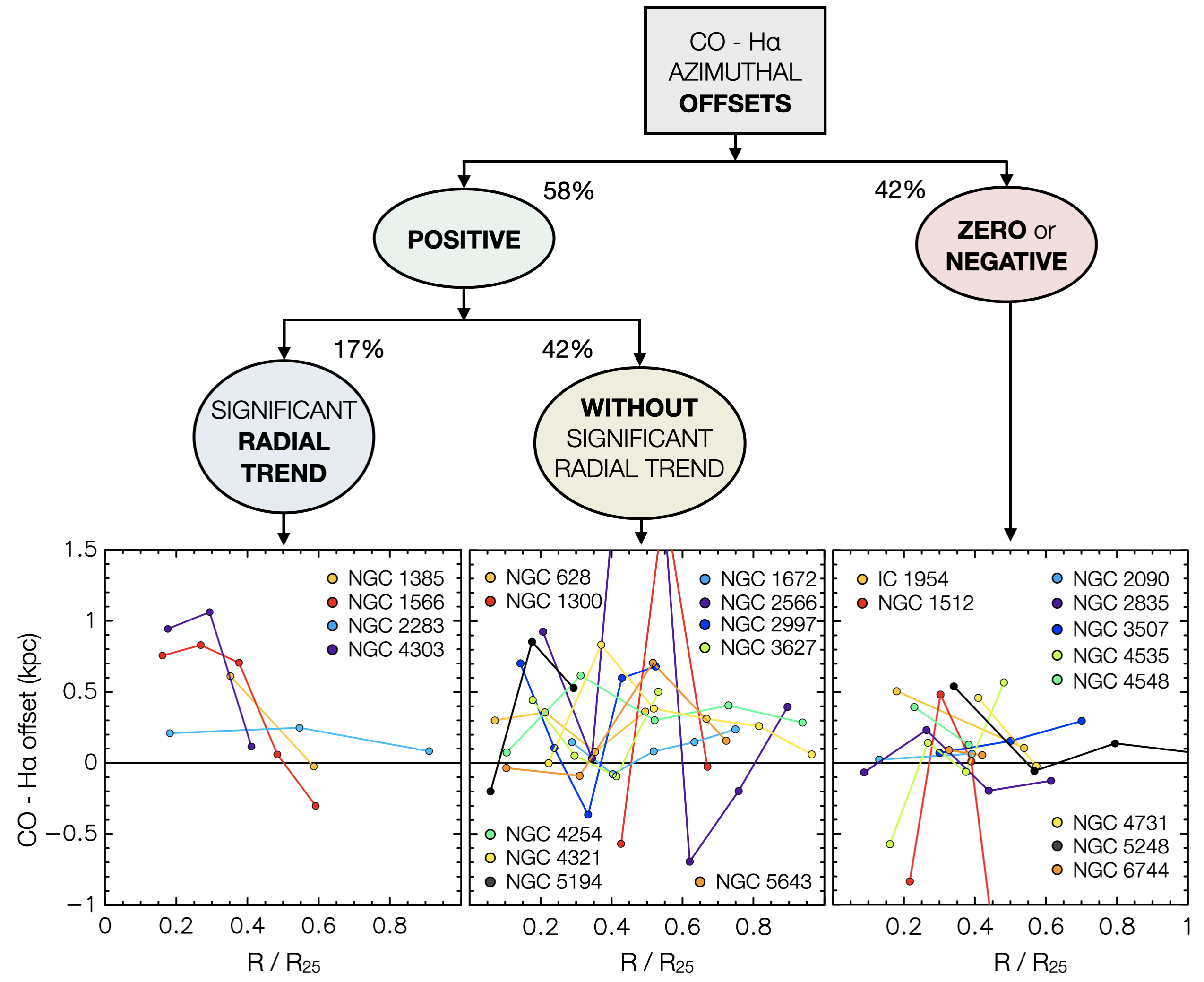}
\end{center}
\caption{Classification of galaxies in our sample into three categories depending on whether they show positive mean CO-H$\alpha$ offsets or not and whether there is a significant trend with radius or not. \mod{The error bars are smaller than the size of the plotted circles.} The solid lines show running means for individual galaxies plotted against radius normalised by $R_{25}$. All galaxies are barred except for NGC\,628, NGC\,1385, NGC\,2090, NGC\,2997, and NGC\,4254.}
\label{fig:classification}
\end{figure*}

Figure~\ref{fig:classification} shows the radial behaviour of mean CO-H$\alpha$ offsets for our sample of spiral galaxies. We split them into three groups, depending on whether they show positive mean offsets and a significant radial trend or not. 58\% of the galaxies show positive offsets. For the remaining 42\% of the galaxies, the offsets are either negative or indistinguishable from zero (we checked if the distribution of offsets within a given galaxy is statistically different from a normal distribution with the same standard deviation randomly distributed around zero offset, with a KS test with a $p$-value less than 0.05 as the mean over 10\,000 runs). These galaxies would be compatible with dynamical (transient) spirals or material arms according to the scenarios outlined in Fig.~\ref{fig:method}.

Among the galaxies with clear positive mean offsets, four galaxies (17\% of the total) also show a significant declining radial trend, with a rank correlation coefficient $\rho < -0.2$ (NGC\,1385, NGC\,1566, NGC\,2283, NGC\,4303). In these cases, the left panel of Fig.~\ref{fig:classification} confirms that the running means generally follow a monotonic declining trend with radius, implying that these galaxies might be compatible with a quasi-stationary spiral density wave according to the expectations from Fig.~\ref{fig:method}. In Sect.~\ref{Sec:OmegaP} below we derive the pattern speed that would be implied for these galaxies following the method from \citet{2009ApJ...697.1870E}.

For the ten remaining galaxies with positive net offsets (42\% of the total), the declining radial correlation is very poor or non-existent ($\rho > -0.2$). This less systematic radial behaviour is also manifested by the oscillations in the running means in the middle panel of Fig.~\ref{fig:classification}. For these galaxies, multiple overlapping modes might be present in the disc. For some galaxies, the offsets at certain radii are exceedingly large because at that point the spiral is nearly tangent to the circular bins where we perform the measurements (very low pitch angle); this is the case for NGC\,1300, for example. In Sect.~\ref{Sec:discussion} we discuss the implications of these findings and caveats in more detail. Fig.~\ref{fig:atlas} shows the individual offsets as a function of radius for each galaxy.

\subsection{Offsets between stellar mass and molecular gas} 
\label{Sec:NIR-CO}

If we follow \citet{1969ApJ...158..123R}, gas in spiral arms is expected to shock immediately before the potential minimum; thus, CO should peak before the maximum stellar mass surface density, leading to positive CO-NIR offsets inside co-rotation (we trace the potential minimum using a NIR stellar mass map; see Sect.~\ref{Sec:stellarmass}). However, simulations with a fixed potential show that the position of the shock compared to the potential minimum actually depends on the gas sound speed \citep{2007PhDT.........1D,2008ApJ...675..188W}. 
In this context, we may expect more mixed results for CO-NIR offsets than for CO-H$\alpha$ even in the context of quasi-stationary density wave theory.

Figure~\ref{fig:histo_CO-Ha} shows that CO-NIR offsets also tend to be predominantly positive, in agreement with the classical picture from \citet{1969ApJ...158..123R}. We also find a net decrease in the mean offset when shifting from inner to outer radii. Thus, the study of CO-NIR offsets paints a similar picture as the study of CO-H$\alpha$ offsets. This connection is also confirmed by the fact that despite significant scatter, the individual CO-H$\alpha$ and CO-NIR offsets display a moderate correlation with each other ($\rho = 0.43$).

\begin{figure*}[t]
\sidecaption
\includegraphics[trim=0 0 0 0, clip,width=12.5cm]{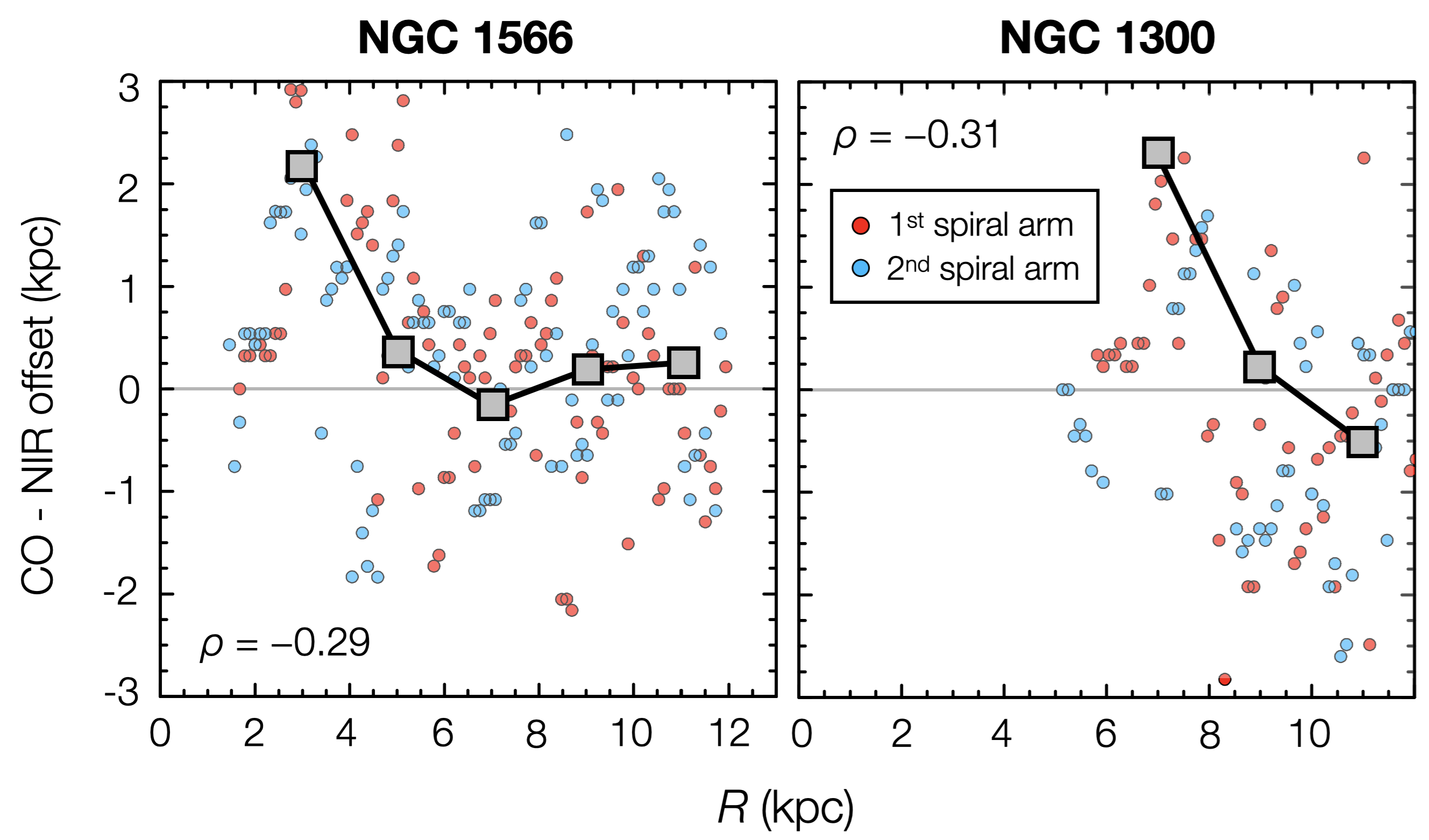}
\caption{CO-NIR offsets as a function of galactocentric radius for the galaxies NGC\,1566 and NGC\,1300. \mod{The error bars are comparable to the size of the plotted circles.} The $\rho$ value in the corner indicates the strength of the correlation (Spearman rank coefficient).}
\label{fig:examples_CO-NIR}
\end{figure*}

Figure~\ref{fig:examples_CO-NIR} shows the distribution of CO-NIR offsets as a function of galactocentric radius in NGC\,1566 and NGC\,1300. This demonstrates that despite a large scatter, there is a mild radial trend. 

Table~\ref{table:stats} lists the average azimuthal offsets between CO and H$\alpha$, as well as between CO and NIR, for each galaxy in our sample. This confirms that despite the bulk mean offsets that we measure, the scatter of individual offset measurements is generally large. The median of the standard deviation in the offsets is ${\sim}1$\,kpc. This makes it clear that additional factors come to play in determining the actual local offset for a given location. We discuss this in more detail in Sect.~\ref{Sec:scatter} below.

Tracing the potential minimum is less straightforward than locating CO or H$\alpha$ peaks (because the stellar mass contrasts are smaller, e.g. \citealt{2024A&A...687A.293Q} and many previous works). To confirm the positive bulk CO-NIR offsets, we repeated the measurements with an alternative tracer, JWST NIRCam F200W instead of the ICA-based stellar mass map. For the galaxies where JWST NIRCam mapping is available, the mean offset is $433$\,pc for ICA and $291$\,pc for NIRCam. It is worth noting that there are significant local differences, though, which make it clear that pinpointing the minimum of the gravitational potential within spiral arms is challenging, but the fact that there is a net positive CO-NIR offset within our field-of-view seems robust.

\begin{table*}[t!]
\begin{center}
\caption[h!]{Mean CO-H$\alpha$ and CO-NIR offsets and their scatter.}
\begin{tabular}{lcccccccccccc}
\hline\hline
\noalign{\smallskip}
&	$\langle \theta_{\rm CO-H\alpha} \rangle$	 	&	$\sigma$	&	$\langle \theta_{\rm CO-H\alpha}^{\rm inner} \rangle$	 	&	$\sigma$	&	$\langle \theta_{\rm CO-H\alpha}^{\rm outer} \rangle$	 	&	$\sigma$	&	$\langle \theta_{\rm CO-NIR} \rangle$	 	&	$\sigma$	&	$\langle \theta_{\rm CO-NIR}^{\rm inner} \rangle$	 	&	$\sigma$	&	$\langle \theta_{\rm CO-NIR}^{\rm outer} \rangle$	 	&	$\sigma$	\\
\noalign{\smallskip}
   \hline
\noalign{\smallskip}
  IC1954 &  206$^\star$ &  1040 &  242$^\star$ &  1029 &  171$^\star$ &  1062 &  536  &  1157 &  536  &  717 &  536$^\star$ &  1469 \\
  NGC0628 &  236  &  948 &  343  &  1223 &  152  &  650 &  220  &  934 &  185$^\star$ &  1138 &  247  &  739 \\
  NGC1300 &  457  &  3587 &  656  &  4318 &  278$^\star$ &  2801 &  526  &  2728 &  1660  &  3506 &  -485$^\star$ &  1021 \\
  NGC1385 &  265  &  1040 &  503  &  1215 &  21.8$^\star$ &  767 &  370$^\star$ &  1272 &  451$^\star$ &  1344 &  288$^\star$ &  1205 \\
  NGC1512 &  -402  &  3083 &  -9.0$^\star$ &  3761 &  -916$^\star$ &  1784 &  990$^\star$ &  4702 &  2142  &  5928 &  -516$^\star$ &  1193 \\
  NGC1566 &  403  &  1064 &  750  &  916 &  60.1  &  1094 &  545  &  1452 &  990  &  1627 &  105$^\star$ &  1097 \\
  NGC1672 &  106  &  582 &  10.0$^\star$ &  789 &  178  &  352 &  184  &  570 &  -67.5$^\star$ &  667 &  372  &  397 \\
  NGC2090 &  32.5$^\star$ &  596 &  7.4$^\star$ &  676 &  56.1$^\star$ &  520 &  32.2$^\star$ &  798 &  93.3$^\star$ &  822 &  -25.1$^\star$ &  783 \\
  NGC2283 &  201  &  752 &  411  &  739 &  -0.2$^\star$ &  716 &  64.6  &  703 &  179$^\star$ &  739 &  -45.5$^\star$ &  656 \\
  NGC2566 &  737  &  5110 &  1509  &  6182 &  78.7  &  3900 &  1206  &  4866 &  2714  &  6153 &  -78.8  &  2878 \\
  NGC2835 &  -11.3  &  852 &  123  &  997 &  -181  &  586 &  -232  &  998 &  -46.7  &  1131 &  -466  &  743 \\
  NGC2997 &  250  &  2135 &  199  &  1244 &  311  &  2876 &  618  &  1857 &  194  &  1052 &  1133  &  2420 \\
  NGC3507 &  146$^\star$ &  1757 &  232$^\star$ &  1622 &  55.6$^\star$ &  1908 &  752$^\star$ &  1920 &  844$^\star$ &  1907 &  655$^\star$ &  1956 \\
  NGC3627 &  176  &  962 &  251  &  827 &  83.7$^\star$ &  1107 &  -154$^\star$ &  1157 &  -141$^\star$ &  1041 &  -170$^\star$ &  1294 \\
  NGC4254 &  377  &  1093 &  357  &  918 &  410  &  1344 &  55.2$^\star$ &  1175 &  251$^\star$ &  988 &  -273$^\star$ &  1382 \\
  NGC4303 &  691  &  2070 &  1181$^\star$ &  2191 &  316$^\star$ &  1922 &  -229$^\star$ &  3911 &  681$^\star$ &  5244 &  -925$^\star$ &  2317 \\
  NGC4321 &  409  &  1738 &  590  &  2018 &  86.1  &  1018 &  903  &  2317 &  1400  &  2599 &  17.4$^\star$ &  1319 \\
  NGC4535 &  0.1  &  937 &  -155$^\star$ &  1070 &  209  &  680 &  269$^\star$ &  1112 &  288$^\star$ &  1062 &  244$^\star$ &  1189 \\
  NGC4548 &  254$^\star$ &  896 &  394$^\star$ &  639 &  128$^\star$ &  1073 &  -332$^\star$ &  1090 &  40.9$^\star$ &  672 &  -665$^\star$ &  1282 \\
  NGC4731 &  188$^\star$ &  575 &  355$^\star$ &  531 &  -10.9$^\star$ &  571 &  193$^\star$ &  965 &  162$^\star$ &  701 &  229$^\star$ &  1221 \\
  NGC5194 &  399  &  1111 &  29.3$^\star$ &  923 &  761  &  1164 &  335  &  761 &  152$^\star$ &  481 &  516  &  927 \\
  NGC5248 &  71.8$^\star$ &  798 &  209  &  891 &  -110$^\star$ &  616 &  -212  &  656 &  -117  &  750 &  -338  &  481 \\
  NGC5643 &  282  &  2114 &  295  &  2481 &  269$^\star$ &  1671 &  265  &  2240 &  160  &  2663 &  373  &  1712 \\
  NGC6744 &  1.2$^\star$ &  1051 &  307$^\star$ &  1388 &  -37.5$^\star$ &  1006 &  457  &  2770 &  -98.9$^\star$ &  912 &  527  &  2918 \\
\noalign{\smallskip}
 \hline
 \hline
\end{tabular}
\label{table:stats}
\end{center}
\tablefoot{Average azimuthal offsets among tracers (CO and H$\alpha$, CO and NIR) for each galaxy considering all radial bins, $\langle \theta_{\rm CO-X} \rangle$, the innermost half of the radial bins, $\langle \theta_{\rm CO-X}^{\rm inner} \rangle$, and the outermost half of the radial bins, $\langle \theta_{\rm CO-X}^{\rm outer} \rangle$. The $\sigma$ column next to each average value shows the standard deviation of that distribution. The $\star$ superscript indicates that the distribution of offsets is statistically indistinguishable from a normal distribution with the same standard deviation randomly distributed around zero (KS test with mean $p$-value greater than 0.05 over 10\,000 runs).}
\end{table*}

\subsection{Deriving star formation timescales and pattern speeds} 
\label{Sec:OmegaP}

\begin{table}[t!]
\begin{center}
\caption[h!]{Pattern speed and star formation timescales.}
\begin{tabular}{llccccc}
\hline\hline 
galaxy	&	arm	&	$\rho$	&	$\Omega_{\rm P}$	&	$\Delta \Omega_{\rm P}$	&	$t_{\rm H\alpha}$	&	$\Delta t_{\rm H\alpha}$	\\
\hline
NGC1385	&	all	&	0.31	&	26.1	&	13.8	&	8.96	&	3.32	\\
NGC1385	&	sp1	&	0.25	&	25.8	&	18.3	&	8.92	&	4.42	\\
NGC1385	&	sp2	&	0.36	&	26.3	&	21.6	&	9.00	&	5.18	\\
NGC1566	&	all	&	0.56	&	25.0	&	4.8	&	2.47	&	0.32	\\
NGC1566	&	sp1	&	0.50	&	24.0	&	6.0	&	2.74	&	0.46	\\
NGC1566	&	sp2	&	0.61	&	26.0	&	7.0	&	1.77	&	0.33	\\
NGC2283	&	all	&	0.32	&	28.2	&	7.0	&	1.79	&	0.29	\\
NGC2283	&	sp1	&	0.31	&	29.9	&	11.4	&	4.72	&	1.20	\\
NGC2283	&	sp2	&	0.37	&	28.0	&	9.3	&	1.18	&	0.25	\\
NGC4303	&	all	&	0.40	&	34.8	&	20.1	&	12.27	&	4.93	\\
NGC4303	&	sp1	&	0.25	&	38.1	&	65.2	&	27.39	&	32.76	\\
NGC4303	&	sp2	&	0.46	&	32.9	&	12.8	&	5.15	&	1.41	\\\hline
 \hline
\end{tabular}
\label{table:OmegaP}
\end{center}
\tablefoot{$\rho$ is the strength of the correlations of $\theta$ (deg) vs $\Omega$ (km\,s$^{-1}$\,kpc$^{-1}$), and the implied pattern speed and star formation timescale following \citet{2009ApJ...697.1870E} for the four galaxies with mean positive offsets and a significant radial trend. `sp1' and `sp2' denote the fits for individual spiral arms, whereas `all' lists the fits for all spiral arms in a given galaxy simultaneously. Pattern speeds are in units of km\,s$^{-1}$\,kpc$^{-1}$ and star formation timescale are in units of Myr.}
\end{table}

For the four galaxies where we found positive mean CO-H$\alpha$ offsets and a significant decreasing radial trend ($\rho < -0.2$), namely NGC\,1385, NGC\,1566, NGC\,2283, and NGC\,4303, we followed \citet{2009ApJ...697.1870E} to infer the pattern speed and star formation timescales that would be implied by the radial trend of the offsets. We fitted the offsets as a function of $\Omega = v_{\rm rot} / R$ according to Fig.~\ref{fig:spiral_sketch} (accounting for the error bars). The resulting fits for the pattern speeds are listed in Table~\ref{table:OmegaP}, including also fits for each individual spiral arms.

For a few galaxies in our sample, \citet{2021AJ....161..185W} determined the pattern speed using the Tremaine-Weinberg (TW) method \citep{1984ApJ...282L...5T}. This method was applied to several tracers (CO, H$\alpha$, and MUSE stellar kinematics), and the conclusion, which agrees with previous expectations, is that the most reliable pattern speeds are those derived applying TW to MUSE stellar kinematics (confirmed by \citealt{2023MNRAS.524.3437B} using simulations).

For NGC\,1385, \citet{2021AJ....161..185W} derived $\Omega_{\rm P} = 25.0^{+5.6}_{-19.3}$\,km\,s$^{-1}$\,kpc$^{-1}$ for the stellar TW, which is in excellent agreement with our derived value of $\Omega_{\rm P} = 26.1 \pm 13.8$\,km\,s$^{-1}$\,kpc$^{-1}$, though the error bars are large in both cases. \mod{Our} value of $\Omega_{\rm P} = 26.1$\,km\,s$^{-1}$\,kpc$^{-1}$ corresponds to a co-rotation radius at $R_{\rm CR}=4.8$\,kpc.

NGC\,1566 is the galaxy that shows the strongest correlation coefficient between CO--H$\alpha$ and $\Omega$ in our sample ($\rho \sim 0.6$). The implied pattern speed, which is in excellent agreement between both spiral arms, is $25.0 \pm 4.8$\,km\,s$^{-1}$\,kpc$^{-1}$ (which corresponds to $R_{\rm CR}=7.6$\,kpc). For NGC\,1566, \citet{2021AJ....161..185W} obtained $\Omega_{\rm P} = 29.4^{+9.1}_{-7.2}$\,km\,s$^{-1}$\,kpc$^{-1}$ for MUSE stellar kinematics, compatible within $1\sigma$ with our value. \citet{1983IAUS..100..151C} derived a slightly lower value of $\Omega_{\rm P} = 18.25$\,km\,s$^{-1}$\,kpc$^{-1}$ based on morphology, while \citet{2005astro.ph..9708K} obtained $\Omega_{\rm P} = 33$\,km\,s$^{-1}$\,kpc$^{-1}$ comparing with tailored hydrodynamical simulations. Thus, our value lies well within the range spanned by the previous determinations from the literature. 

For NGC\,2283, \citet{2021AJ....161..185W} only applied TW to the molecular gas maps (because no MUSE data were available for this galaxy), obtaining $\Omega_{\rm P} = 41.0^{+1.9}_{-3.1}$\,km\,s$^{-1}$\,kpc$^{-1}$. This is significantly larger than our derived value of $\Omega_{\rm P} = 28.2 \pm 7.0$\,km\,s$^{-1}$\,kpc$^{-1}$; however, as commented above, applying TW to the clumpy gas component cannot be considered as reliable as the measurements based on stellar kinematics \citep{2021AJ....161..185W,2023MNRAS.524.3437B}.

For NGC\,4303, \citet{2021AJ....161..185W} obtained $\Omega_{\rm P} = 43.5^{+5.3}_{-10.0}$\,km\,s$^{-1}$\,kpc$^{-1}$ for MUSE stellar kinematics, which is slightly larger than our determination of $\Omega_{\rm P} = 34.8 \pm 20.1$\,km\,s$^{-1}$\,kpc$^{-1}$ but agrees within $1\sigma$ if we consider the significant error bar in this case. For NGC\,4303 (as well as for NGC\,1385 and NGC\,1566), the field of view of the MUSE observations used by  \citet{2021AJ....161..185W} was very similar to the field of view of the PHANGS-ALMA maps used here, which means that both studies focused on the same radial range within the galaxies. Our pattern speed for NGC\,4303 implies a co-rotation radius at $R_{\rm CR}=5.9$\,kpc. The fact that this is well outside the bar co-rotation radius ($R_{\rm CR}^{\rm bar}=3.4 \pm 0.2$\,kpc; \citealt{2024A&A...691A.351R}) confirms that the bar and spiral do not rotate with the same pattern speed (i.e.\ they are not directly coupled).

Table~\ref{table:OmegaP} also lists the star formation timescales derived fitting the CO-H$\alpha$ offsets as a function of $\Omega$. They are typically of several megayears (ranging $2{-}12$\,Myr for the different galaxies), which broadly agrees with the star formation timescale of a few megayears associated with H$\alpha$ as measured for PHANGS galaxies (e.g.\ \citealt{2020MNRAS.493.2872C,2021MNRAS.504..487K,2022MNRAS.516.3006K}).

\section{Discussion}
\label{Sec:discussion}

\subsection{Do stars and gas cross spiral arms?} 
\label{Sec:discussion_cross}

As explained in the introduction, theories on the nature of spiral arms can be broadly divided into two main classes. The first class posits that spiral arms are always made up of the same material; they can either be dynamical spirals (transient features which might be self-excited and swing-amplified), or material arms, which are relatively long-lived (possibly tidally induced by an interaction, e.g.\ \citealt{1969ApJ...158..899T,1981seng.proc..111T}). A second class encompasses quasi-stationary density waves and, in theory, also kinematic spirals \citep{1973PASA....2..174K}, where stars and gas flow through the spiral arms \citep{2014PASA...31...35D}. As explained in Sect.~\ref{Sec:method}, offsets can allow us to differentiate between these two major classes of spiral theories.

We examined a \mod{set} of well-delineated spirals from the PHANGS sample. Despite significant scatter, in 58\% of the galaxies we find net positive offsets between CO and H$\alpha$, suggesting that the gas rotates faster than the spiral perturbation (i.e.\ gas `overtakes' the spiral). This explains why, on average, young stars associated with H$\alpha$ are displaced towards the leading side of the spiral (i.e.\ `downstream') compared to the maximum accumulation of molecular gas traced by CO. Below in Sect.~\ref{Sec:scatter} we discuss the implications of the scatter in the offsets.

If we restrict ourselves to the few scenarios introduced in Fig.~\ref{fig:method}, our observations would support some form of rigidly rotating (though not necessarily long-lived) density wave modes for nearly two-thirds of the sample. The net positive offsets that we measure per galaxy are typically ${\sim}200{-}300$\,pc in the plane of the galaxy. This is consistent with the displacement expected within a timescale of a few megayears. Indeed, for example, for a radius of $R \sim 5$\,kpc, a representative rotation curve of $v_{\rm rot} \sim 150$\,km~s$^{-1}$ \citep{2020ApJ...897..122L}, and a reasonable pattern speed of $\Omega_{\rm P} \sim 20$\,km~s$^{-1}$~kpc$^{-1}$, an offset of $200{-}300$\,pc corresponds to a timescale of ${\sim}4{-}6$\,Myr, which matches the star formation timescales inferred for PHANGS galaxies using H$\alpha$ as a star formation tracer (e.g.\ \citealt{2020MNRAS.493.2872C,2021MNRAS.504..487K,2022MNRAS.516.3006K}). Therefore, a priori, the magnitude of the offsets seems roughly consistent with the expectation for a representative rotation curve and a reasonable spiral pattern speed. Yet, it is important to emphasise that while we use gas and young stars as tracers, this does not imply that a density wave is sustained in the gaseous disc (which is sensitive to complex hydrodynamical and feedback effects); gas might just be locally responding to an underlying spiral density wave present in the old stellar population (which would agree with the modest but clearly positive $\Sigma_{\star}$ arm/interarm contrasts found by \citealt{2024A&A...687A.293Q}). \mod{The typical mean offset of $200{-}300$\,pc per galaxy corresponds to an angular offset of $5{-}7\arcsec$ for our nearest targets ($d \sim 9$\,Mpc) and $2{-}3\arcsec$ for the most distant ones ($d \sim 23$\,Mpc); this could explain why some previous studies with coarser resolution did not detect significant offsets, especially for the most distant galaxies.}

For 42\% of the galaxies, we find CO-H$\alpha$ offsets which are indistinguishable from zero or even negative. In these cases, it is possible that gas and stars are actually co-rotating with the spiral perturbation, without crossing the arms. At least two mechanisms can explain this scenario: dynamic spirals which can appear in isolated discs and material arms which can be tidally excited by the passage of a companion galaxy. The former consist of local hydrodynamical perturbations which are swing-amplified and sheared into spiral segments as a consequence of differential rotation; these are commonly found in hydrodynamical simulations \citep[e.g.][]{1984ApJ...282...61S,2014ApJ...785..137S,2013ApJ...766...34D,2015MNRAS.449.3911P,2020MNRAS.498.1159P,2024MNRAS.532..126D}.
On the other hand, material arms can be long and coherent, potentially tidally dragged by a companion during an interaction. The fact that these galaxies show positive $\Sigma_{\star}$ arm/interarm contrasts \citep{2024A&A...687A.293Q} would argue in favour of material arms as opposed to shorter transient spirals, which would not necessarily stand out in the old stellar component.

In principle, if we sample well beyond co-rotation a galaxy with a single quasi-stationary spiral density wave, the average offset could be zero (because positive offsets inside co-rotation could compensate negative offsets outside). However, this possibility is ruled out by Fig.~\ref{fig:classification} because we did not find any galaxies \mod{with zero mean offsets} where running means monotonically decrease from positive to negative as radius increases. In any case, the limited PHANGS-ALMA fields of view might be obscuring this (see Sect.~\ref{Sec:caveats}).

Following the spiral family nomenclature from \citet{1990NYASA.596...40E}, our galaxy sample is made up of 15 grand-design, seven multi-armed, and two flocculent spirals (following \citealt{2015ApJS..217...32B} and \citealt{2021ApJ...913..113M}, as listed in \citealt{2024A&A...687A.293Q}; see Table~\ref{table:sample}). One could perhaps expect a higher proportion of grand-design spirals within the group of galaxies with positive offsets and a clear trend, compatible with a quasi-stationary density wave. Interestingly, this is not the case, and among the four galaxies in that group, only one is grand-design; perhaps even more surprisingly, both flocculent spirals (NGC\,1385 and NGC\,2283) fall into this category. This questions the widespread idea that grand-design spirals are generally associated with quasi-stationary density waves while flocculent spirals are made up of transient, dynamic spirals \citep{2014PASA...31...35D}. It is worth noting that none of the galaxies in this first group seems to be interacting; galaxies with clear signs of interaction either show positive offsets without a clear radial trend, which might be evidence of multiple modes (e.g.\ NGC\,3627, NGC\,4321, NGC\,5194), or they do not show net positive offsets at all, which would agree with material arms driven by the interaction (NGC\,1512, NGC\,4731). \mod{In any case, none of these galaxies are evolved ongoing mergers comparable to the Antennae; they are interacting pairs, typically separated by distances of ${\sim}100$\,kpc or more. The only case of a closely interacting pair is M51 (NGC\,5194), and, to a lesser extent, NGC\,1512.}

In addition to galaxy interactions, bars have also been argued to have the ability to induce spiral arms \citep[e.g.][]{1976ApJ...209...53S,1988MNRAS.231P..25S,2004AJ....128..183B,2010MNRAS.407.1433A,2010ApJ...715L..56S}.
Our sample is overall clearly biased towards barred galaxies, since as many as 19 out of 24 galaxies (79\%) are barred according to \citet{2021A&A...656A.133Q}. Except for NGC\,1385, all four galaxies with positive offsets and a clear radial trend have a large-scale bar. In the remaining categories of mean positive offsets without a clear radial trend and zero or negative mean offsets, we find a fraction of $70$\% and $90$\% barred galaxies. Thus, the presence of a bar alone cannot explain the differences we found.

In some grand-design spiral galaxies it has been suggested that a certain radial range behaves as a quasi-stationary density wave, while at other radii it can have a material nature \citep[e.g. in M51;][]{2014ApJ...784....4C,2017ApJ...845...78C}.
The same might apply to other interacting galaxies, some of which are clearly asymmetric (e.g.\ NGC\,3627). As we are mostly sampling the inner part of galaxies, if the inner part behaves as quasi-stationary density wave and the outer one as material arms (as often argued for M51), we would be exclusively sensitive to the inner behaviour. Thus, our results cannot be necessarily extrapolated to the entire galaxy. Also, in principle interactions could induce asymmetries, which may result in a given arm behaving differently from another arm. For this reason, we examine the radial trend of the offsets for each spiral arm independently. Indeed, \citet{2009ApJ...697.1870E} also found differences among spiral arms within certain galaxies. Inspecting Fig.~\ref{fig:atlas} in the Appendix, as well as the corresponding correlation coefficients, we conclude that in most cases the behaviour of one arm is closely mirrored by the second or third arm.

Spiral arms are often recognised as a source of large-scale azimuthal metallicity variations. Some studies identified significant metallicity differences in spiral arms \citep[e.g.][]{2017ApJ...846...39H,2018A&A...618A..64H}, and even more so for grand-design spirals \citep{2020MNRAS.492.4149S}; yet, other studies did not find clear trends associated with spiral arms \citep{2022MNRAS.509.1303W} or did so only for some galaxies \citep{2019ApJ...887...80K}. In this context, recent numerical simulations point to the prerequisite of a radial metallicity gradient, which can then lead to azimuthal metallicity variations as a result of radial gas flows driven by spiral arms \citep{2023A&A...671A..56K,2023MNRAS.521.3708O}. \citet{2024MNRAS.534..883C} examined gas-phase metallicity variations in nine spiral galaxies, with strong evidence supporting quasi-stationary density wave in NGC\,1566 (and not in NGC\,2835 or NGC\,6744), in agreement with our findings. While no correlations between our results and the azimuthal patterns presented in eight galaxies in \citet{2019ApJ...887...80K} are apparent, detailed constraints on the different spiral mechanisms acting in specific galaxies could provide a novel constraint on models, and may aid our understanding of the role spiral arms play in mixing the ISM and driving patterns in the gas-phase abundances.

\subsection{Star formation is not uniquely initiated at a spiral shock} 
\label{Sec:scatter}

As explained in Sect.~\ref{Sec:method}, previous observations demonstrated that star formation in spiral galaxies is not exclusively triggered at the spiral shock. As expected, this results in a large scatter among the azimuthal offsets. While the mean offsets that we measure in our sample are typically of a few hundred parsecs, the scatter of the offsets is of the order of ${\sim}1$\,kpc (Table~\ref{table:stats}).

The large scatter is partially a result of the high resolution of our observations, which reflects local conditions and small-scale physics effects. With our improved resolution compared to previous studies, we start resolving individual clouds and star forming complexes, at scales where star formation is subject to stochastic sampling effects \citep[e.g.][]{2010ApJ...722.1699S}. Gravitational instabilities and local turbulence triggered by supernova bubbles (and other forms of stellar feedback) also play a role in shaping the interstellar medium on small scales, and it can regulate its ability to produce new stars. Specifically, we know that much star formation in spiral arms is not confined to a spiral shock, but taking place in spur- and feather-like gas structures \citep[e.g.][]{2006ApJ...650..818L,2017ApJ...836...62S,2022ApJ...941L..27W}. 

More generally, star formation is not limited to spiral regions, and in absolute terms (total SFR, not $\Sigma_{\rm SFR}$) almost as much star formation takes place in interarm regions as in spiral arms \citep{2010ApJ...725..534F,2024A&A...687A.293Q}.  
This means that mechanisms other than only spiral arms must initiate the collapse of molecular clouds to form stars.
Thus at any location relative to the arms, there will be a mixture of gas and star formation related to both processes, and this naturally injects scatter into the offset measurement. 
In this context, the scatter in our measurements implies that star formation is not uniquely initiated at a coherent spiral shock.

\subsection{Are observations consistent with a single pattern speed?}
\label{Sec:single_OmegaP}

In the framework of quasi-stationary density wave theory, if the entire spiral perturbation rotates as a rigid body (with a single pattern speed), according to Fig.~\ref{fig:method} we would expect mean positive offsets in the inner regions of the galaxy and a declining radial trend. The confirmation that a single pattern speed applies throughout the disc would be given by the observation of negative offsets after the co-rotation radius. This condition including negative offsets after a certain radius is only met by NGC\,1566. In three more galaxies (NGC\,1385, NGC\,2283, and NGC\,4303), we find a significant decreasing radial trend ($\rho < -0.2$), but the running means do not become negative (Fig.~\ref{fig:classification}), probably because the PHANGS-ALMA field of view does not cover sufficiently large radii, i.e.\ it likely samples close to, but not beyond the co-rotation radius of the spiral (as shown in Fig.~\ref{fig:classification} with radii normalised by $R_{25}$). Yet, if we sampled further out, offsets might become larger again, as a consequence of multiple modes dominating at different radial ranges (predicted for groove modes and in the theory of mode coupling; e.g.\ \citealt{1988MNRAS.232..733S,1997A&A...322..442M,2014ApJ...785..137S}). Therefore, at least one and up to four galaxies in our sample seem compatible with a single pattern speed. 

Another point that adds confidence to these results is the fact that the pattern speeds that we derived in Sect.\,\ref{Sec:OmegaP} for the four galaxies with a clear radial trend agree well with previous measurements from the literature. These previous studies relied on totally independent methods such as Tremaine-Weinberg applied on stellar kinematics from MUSE \citep{2021AJ....161..185W}. The star-formation timescales implied by the fitting of the offsets as a function of $\Omega$ following \citet{2009ApJ...697.1870E} are of the order of several megayears, which also roughly agrees with independent empirical estimates (e.g.\ \citealt{2020MNRAS.493.2872C,2021MNRAS.504..487K,2022MNRAS.516.3006K}).

For galaxies with positive mean offsets but no clear overall radial trend, several pattern speeds might be at play in the spiral structure. This is expected in the context of the theory of mode coupling \citep[e.g.][]{1988MNRAS.232..733S,1997A&A...322..442M}.
Simulations such as \citet{1999A&A...348..737R} and \citet{2012MNRAS.426.2089R} found evidence of multiple pattern speeds in simulations of isolated spiral galaxies. Empirically, multiple pattern speeds have been identified for some galaxies, for example by \citet{2008ApJ...688..224M,2009ApJ...702..277M} and \citet{2014ApJS..210....2F}.
\citet{2024RAA....24g5007K} compiled measurements of co-rotation radii from the literature for hundreds of galaxies, based on different methods, and found them to be inconsistent for most galaxies and interpret this as evidence of either multiple spiral modes rotating at different angular velocities or transient spirals.
Our sample does not have any galaxies in common with \citet{2009ApJ...702..277M}, but \citet{2008ApJ...688..224M} and \citet{2024ApJ...966..110F} found evidence of several spiral modes in M51 (NGC\,5194), which would agree with the fact that we find a net positive CO-H$\alpha$ offset but no consistent radial trend in this galaxy. Several previous studies did not find evidence supporting the (quasi-stationary) density wave nature of the grand-design spiral NGC\,4321 \citep{2009ApJ...697.1870E,2012MNRAS.424.1636F,2025Galax..13...27K}; in this case, our analysis of CO-H$\alpha$ offsets points to the existence of multiple modes.
The only galaxy in common between our sample and \citet{2014ApJS..210....2F} is NGC\,4303, for which they found as many as six co-rotation radii (including both bar and spiral); the strongest peak in the Font-Beckman method is located at $R_{\rm CR} = 49.9\arcsec$, corresponding to $\Omega_{\rm P} = 39$\,km\,s$^{-1}$\,kpc$^{-1}$, which is well compatible with our result of $\Omega_{\rm P} = 34.8$\,km\,s$^{-1}$\,kpc$^{-1}$ based on the fitting of CO-H$\alpha$ offsets.

\subsection{Robustness of results and caveats} 
\label{Sec:caveats}

Our offset measurements rely on galactocentric radial bins, which implicitly assume circular trajectories for gas and stars. However, stellar orbits in a real non-axisymmetric two-armed spiral potential are rather elliptical and elongated in the direction of the $m=2$ perturbation at each radius (therefore, precessing with radius). This means that if there is a systematic displacement between gas and star formation \mod{along the orbit}, that distance will be underestimated by our circular bins, especially for more open spirals (large pitch angle) with strong stellar arm/interarm contrasts. However, this does not change the fact that offsets should be positive inside co-rotation and negative outside, flipping sign precisely at co-rotation. Thus, the simple picture that we have painted in the previous sections should remain qualitatively valid, but the detailed numerical results could be affected by this limitation. In terms of fitting pattern speeds following \citet{2009ApJ...697.1870E}, we expect the choice of circular sampling to have a stronger impact on the derivation of star formation timescales (which depend on the slope of the fitted line) than the pattern speed (which depends on the intercept), as long as our measurements have a good sampling around co-rotation, where the offsets are expected to be zero and flip their sign.

Since we focus on offsets in the plane of the galaxy, our measurements will be affected by errors in the determination of the inclination and position angle of the disc. Fortunately, for PHANGS galaxies we count on reliable kinematic estimates of these orientation parameters \citep{2020ApJ...897..122L}. We find that perturbing the orientation parameters within their error bars has a typical impact on the offsets and correlation coefficients of at most a few multiples of 10\%, and importantly, it does not significantly wash out any radial trends.

Following \citet{2009ApJ...697.1870E}, we measured the azimuthal offsets between CO and H$\alpha$ peaks at each radius. In order to assess the robustness of this approach, in Appendix~\ref{sec:appendix_robustness} we also considered the impact of choosing the second brightest azimuthal bin in CO and H$\alpha$ at each radius or the mean position weighted by CO or H$\alpha$ intensity. While individual offset measurements change significantly, our conclusions are not qualitatively affected when adopting either of these alternatives. We also checked how adopting wider radial bins ($\Delta R = 300$\,pc) affects the offsets and correlation coefficients with radius, but reassuringly, these only change at the level of a few percent.

Another methodological detail that might be questionable is the choice of H$\alpha$ to trace recent star formation, as this optical tracer is known to suffer from severe extinction. While this is true, we are interested in a tracer that emerges after a delay following the onset of star formation, i.e.\ we aim to `clock' the time between the collapse of molecular gas and the appearance of young exposed stars that emit in H$\alpha$, which makes this an appropriate tracer. \mod{However, dust extinction could be stronger in `upstream' locations compared to `downstream' regions (due to higher gas column densities), and it could also be stronger near the centres of the galaxies. This could lead to stronger CO-H$\alpha$ offsets as a function of radius. Yet, as the contribution of these effects should vary monotonically with radius, we would not expect this caveat to limit our ability to distinguish between galaxies with globally positive or zero CO-H$\alpha$ offsets, and it should not prevent us from identifying present radial trends (but the slope of the radial trend would be affected, and therefore the implied star formation timescales could certainly be over- or underestimated).}

\mod{Another related} problem could be associated with changing star formation timescales \citep[see e.g.][]{2024ARA&A..62..369S}, which could make the radial behaviour more erratic. Specifically, density and other factors affect the time for a star-forming cloud to collapse and form stars \citep[e.g.][]{2022AJ....164...43S,2023MNRAS.521.3348N}. In particular, \citet{2022MNRAS.516.3006K} found star formation timescales associated with H$\alpha$ that vary in the range from ${\sim}5$\,Myr to ${\sim}10$\,Myr for PHANGS galaxies.

It is worth noting that even in the framework of quasi-stationary density wave theory, the mean measured offset for a given galaxy will depend on the field of view. If our maps only sample the innermost regions, the mean offset will be larger, as the difference between $v_{\rm gas} = \Omega_{\rm gas}  R$ and $v_{\rm spiral} = \Omega_{\rm P}  R$ increases with decreasing radius. This can explain why the mean offset is so large for galaxies such as NGC\,2566 ($\langle \theta_{\rm CO-H\alpha} \rangle = 740$\,pc), where our ALMA map samples the whole bar but only the beginning of the spiral structure. Fig.~\ref{fig:classification} allows one to assess this limitation visually, as it displays the running means as a function of $R/R_{25}$; in most galaxies we sampled out to ${\sim}0.5R_{25}$, with a few CO maps extending further out (${\sim}0.8{-}0.9R_{25}$).
The limited field of view also affects our ability to test whether offsets indeed become negative after a certain radius, as we commented on in Sect.~\ref{Sec:single_OmegaP}. This emphasises the importance of wide-field CO mapping for studies such as this one.

\section{Summary and conclusions} 
\label{Sec:concl}

Spiral arms are some of the most compelling features in nearby disc galaxies, but from a dynamical point of view, their nature remains an open question. Azimuthal offsets between molecular gas and star formation can inform us about the nature of the underlying spiral perturbation. Here we have examined a sample of 24 spiral galaxies from the PHANGS survey with a well-delineated spiral structure, quantifying the offsets between CO and H$\alpha$ peaks within spiral regions in radial bins. Our main findings are as follows:

\begin{enumerate}

\item Offsets exhibit substantial scatter, supporting previous findings that star formation is not exclusively initiated at the spiral shock (Table~\ref{table:stats}). Star formation off spiral arms, gravitational instabilities giving rise to spurs and feathers, and turbulence and stellar feedback likely play roles in setting this large scatter among the measured offsets.

\item Most galaxies exhibit positive CO-H$\alpha$ and CO-NIR offsets, typically of a few hundred parsecs (Fig.~\ref{fig:histo_CO-Ha}). This corresponds to a star formation timescale of several megayears for a characteristic radius, rotation curve, and spiral pattern speed, which is consistent with previous observations.

\item In four galaxies (17\%), we find positive mean offsets and a clear radial trend (Fig.~\ref{fig:classification}). This agrees with the expectation for a quasi-stationary spiral density wave with a single pattern speed. For ten galaxies (42\%), the mean offset is positive, but there is not such a clear radial trend, which could suggest the superposition of several modes (such that a spiral mode might not develop beyond its own co-rotation). Yet, we caution that field of view limitations complicate the interpretation of these measurements.

\item In the remaining ten galaxies (42\%), we did not find significant positive offsets (Fig.~\ref{fig:classification}). This would be compatible with transient, dynamical spirals and with material arms possibly dragged by the interaction with a companion. Under either of these scenarios, gas and stars do not overtake the spiral perturbation (i.e.\ all radii are co-rotating).

\item For the four galaxies with clear offsets and a radial trend, we derived pattern speeds in good agreement with the literature (Table~\ref{table:OmegaP}).

\end{enumerate} 

In summary, our results suggest that even well-delineated spirals in the local Universe exhibit significantly different behaviours in terms of CO-H$\alpha$ offsets and their radial distribution. This points to the idea that different kinds of mechanisms are probably responsible for sustaining their spiral structure.

\small  
%
\begin{acknowledgements}   
This work was carried out as part of the PHANGS collaboration. 
\mod{We would like to thank the anonymous referee for constructive comments that helped us improve the manuscript.}
MQ, SGB, MRG, MJJD and AU acknowledge support from the Spanish grant PID2022-138560NB-I00, funded by MCIN/AEI/10.13039/501100011033/FEDER, EU.
HAP acknowledges support from the National Science and Technology Council of Taiwan under grant 113-2112-M-032-014-MY3.
KK acknowledges support from the Deutsche Forschungsgemeinschaft (DFG, German Research Foundation) in the form of an Emmy Noether Research Group (grant number KR4598/2-1, PI Kreckel) and the European Research Council’s starting grant ERC StG-101077573 (“ISM-METALS"). 
FP acknowledges support from the Horizon Europe research and innovation programme under the Maria Skłodowska-Curie grant “TraNSLate” No 101108180, and from the Agencia Estatal de Investigación del Ministerio de Ciencia e Innovación (MCIN/AEI/10.13039/501100011033) under grant (PID2021-128131NB-I00) and the European Regional Development Fund (ERDF) ``A way of making Europe''. 
MQ acknowledges discussions with John Beckman and Sim Dlamini.
MB acknowledges support from the ANID BASAL project FB210003. This work was supported by the French government through the France 2030 investment plan managed by the National Research Agency (ANR), as part of the Initiative of Excellence of Université Côte d’Azur under reference number ANR-15-IDEX-01.
RSK acknowledges financial support from the ERC via Synergy Grant ``ECOGAL'' (project ID 855130),  from the German Excellence Strategy via the Heidelberg Cluster ``STRUCTURES'' (EXC 2181 - 390900948), and from the German Ministry for Economic Affairs and Climate Action in project ``MAINN'' (funding ID 50OO2206).  RSK also thanks the 2024/25 Class of Harvard Radcliffe Fellows for highly interesting and stimulating discussions. 
This work is based on observations and archival data obtained with the \textit{Spitzer} Space Telescope, which is operated by the Jet Propulsion Laboratory, California Institute of Technology under a contract with NASA.
This paper makes use of the following ALMA data:  ADS/JAO.ALMA\#2012.1.00650.S,  
ADS/JAO.ALMA\#2013.1.00803.S,  
ADS/JAO.ALMA\#2013.1.01161.S,  
ADS/JAO.ALMA\#2015.1.00121.S,  
ADS/JAO.ALMA\#2015.1.00782.S,  
ADS/JAO.ALMA\#2015.1.00925.S,  
ADS/JAO.ALMA\#2015.1.00956.S,  
ADS/JAO.ALMA\#2016.1.00386.S, 
ADS/JAO.ALMA\#2017.1.00886.L,  
ADS/JAO.ALMA\#2018.1.01651.S.  
ALMA is a partnership of ESO (representing its member states), NSF (USA) and NINS (Japan), together with NRC (Canada), MOST and ASIAA (Taiwan), and KASI (Republic of Korea), in cooperation with the Republic of Chile. The Joint ALMA Observatory is operated by ESO, AUI/NRAO and NAOJ. The National Radio Astronomy Observatory is a facility of the National Science Foundation operated under cooperative agreement by Associated Universities, Inc.

\end{acknowledgements}

\bibliography{mq}{}
\bibliographystyle{aa}{}

\newpage

\normalsize

\onecolumn

\begin{appendix}

\section{Spiral galaxy sample}
\label{sec:appendix_sample}

\begin{center}
\begin{longtable}{lccccccc}
\caption{Spiral galaxy sample studied in this paper. \label{table:sample}}\\ 
\hline\hline 
& log($M_\star$/$M_\odot$) & log(SFR/[$M_\odot$\,yr$^{-1}$]) & $R_{\rm 25}$ ($\arcsec$) & $d$ (Mpc) & spiral morphology & bar & class \\ 
\hline
IC\,1954	&	9.7	&	-0.44	&	89.8	&	12.8	&	M	&	1	&	C	\\
NGC\,0628	&	10.3	&	0.24	&	296.6	&	9.8	&	M	&	0	&	B	\\
NGC\,1300	&	10.6	&	0.07	&	178.3	&	19.0	&	G	&	1	&	B	\\
NGC\,1385	&	10.0	&	0.32	&	102.1	&	17.2	&	F	&	0	&	A	\\
NGC\,1512	&	10.7	&	0.11	&	253.0	&	18.8	&	G	&	1	&	C	\\
NGC\,1566	&	10.8	&	0.66	&	216.8	&	17.7	&	G	&	1	&	A	\\
NGC\,1672	&	10.7	&	0.88	&	184.6	&	19.4	&	G	&	1	&	B	\\
NGC\,2090	&	10.0	&	-0.39	&	134.6	&	11.8	&	G	&	0	&	C	\\
NGC\,2283	&	9.9	&	-0.28	&	82.8	&	13.7	&	F	&	1	&	A	\\
NGC\,2566	&	10.7	&	0.94	&	127.7	&	23.4	&	G	&	1	&	B	\\
NGC\,2835	&	10.0	&	0.09	&	192.4	&	12.2	&	G	&	1	&	C	\\
NGC\,2997	&	10.7	&	0.64	&	307.7	&	14.1	&	G	&	0	&	B	\\
NGC\,3507	&	10.4	&	0.00	&	87.5	&	23.5	&	G	&	1	&	C	\\
NGC\,3627	&	10.8	&	0.58	&	308.4	&	11.3	&	G	&	1	&	B	\\
NGC\,4254	&	10.4	&	0.49	&	151.1	&	13.1	&	M	&	0	&	B	\\
NGC\,4303	&	10.5	&	0.73	&	206.6	&	17.0	&	M	&	1	&	A	\\
NGC\,4321	&	10.7	&	0.55	&	182.9	&	15.2	&	G	&	1	&	B	\\
NGC\,4535	&	10.5	&	0.33	&	244.4	&	15.8	&	M	&	1	&	C	\\
NGC\,4548	&	10.7	&	-0.28	&	166.4	&	16.2	&	G	&	1	&	C	\\
NGC\,4731	&	9.5	&	-0.22	&	189.7	&	13.3	&	G	&	1	&	C	\\
NGC\,5194	&	10.6	&	0.64	&	411.3	&	8.6	&	G	&	1	&	B	\\
NGC\,5248	&	10.4	&	0.36	&	122.2	&	14.9	&	G	&	1	&	C	\\
NGC\,5643	&	10.3	&	0.41	&	157.4	&	12.7	&	M	&	1	&	B	\\
NGC\,6744	&	10.7	&	0.38	&	470.0	&	9.4	&	M	&	1	&	C	\\
\hline
\end{longtable}
\tablefoot{\mod{Stellar mass and star formation rates are adopted from \citet{2021ApJS..257...43L} who calculated them using the prescriptions and data from \citet{2019ApJS..244...24L}. These are based on GALEX and WISE radial profiles, translated to physical quantities using recipes (SFR conversion and mass-to-light ratio) designed to match the SED fitting results of \citet{2016ApJS..227....2S,2018ApJ...859...11S}.}
The optical isophotal radius $R_{\rm 25}$ is adopted from RC3 \citep{1991rc3..book.....D} via LEDA, \mod{and distances come from \citet{2021MNRAS.501.3621A}}.
The spiral morphology lists the spiral family according to \citet{2015ApJS..217...32B} or the definitions adopted in \citet{2021ApJ...913..113M} for galaxies outside the S$^4$G survey (`G' stands for grand-design, `M' for multi-armed, and `F' for flocculent). The bar presence (1 for barred) reflects the environmental masks from \citet{2021A&A...656A.133Q}. The last column indicates the classification according to Fig.~\ref{fig:classification}: class A implies positive mean offsets and a significant radial trend; class B corresponds to mean positive offsets but no radial trend; class C implies mean offsets compatible with zero or negative.}
\end{center}

\twocolumn

\section{Assessing the robustness of the offset measurements}
\label{sec:appendix_robustness}

In order to study the robustness of our results, Table~\ref{table:alternative_offsets} shows the impact of two alternative definitions of offsets. Our nominal strategy is to measure offsets in galactocentric radial bins based on the peak emission in CO and H$\alpha$ within each spiral arm, as illustrated by Fig.~\ref{fig:method}. This mimics the definition adopted by \citet{2009ApJ...697.1870E}. However, for a number of reasons discussed in Sect.~\ref{Sec:scatter}, these offsets are subject to very large scatter. Thus, here we consider the impact of choosing the second brightest azimuthal bin at each radius (in CO and H$\alpha$, `2$^{\rm nd}$ peak') instead of the actual peak, to measure how sensitive our results are to the choice of individual peaks. We also consider the weighted mean azimuth (weighted by CO and H$\alpha$ luminosity) for each spiral at each radius. This is expected to mitigate stochasticity associated with individual clumps at the expense of somewhat washing out the signal due to emission that is more spread out and not directly associated by the accumulation of CO due to the spiral or HII regions which are the direct product of spiral arm star formation.

\begin{figure}[t]
\begin{center}
\includegraphics[trim=0 0 0 0, clip,width=0.45\textwidth]{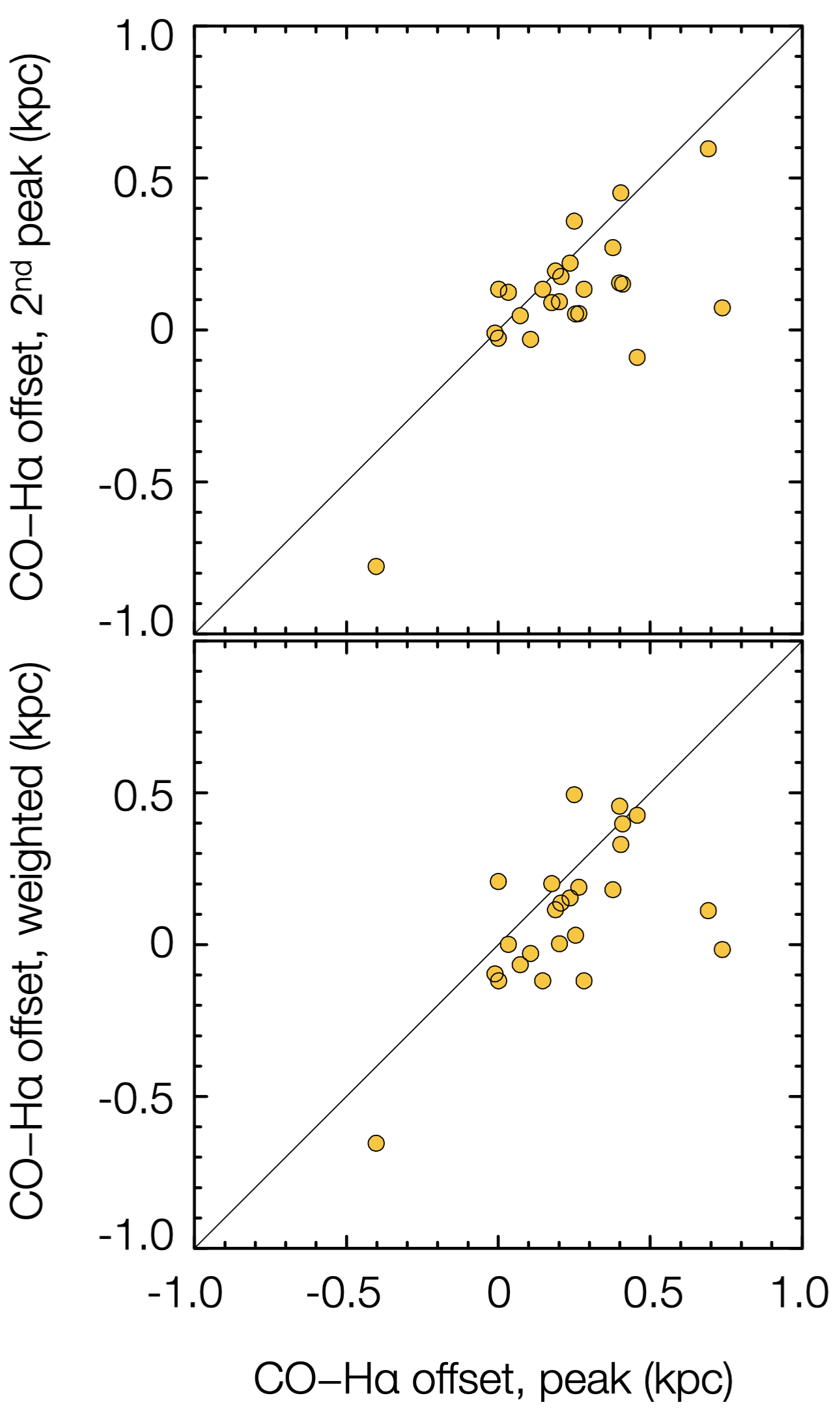}
\end{center}
\caption{\mod{Mean CO-H$\alpha$ offsets based on our nominal approach (intensity peaks) compared to two alternative definitions of offsets: choosing the second brightest peak or the intensity-weighted position (see text for details). Each data point represents the mean offset per galaxy.}}
\label{fig:appendix_peak}
\end{figure}

The results from Table~\ref{table:alternative_offsets} \mod{and Fig.~\ref{fig:appendix_peak}} confirm that the choice of the second brightest azimuthal bin or the weighted mean position does have a significant impact on the individual offsets. Yet, the main conclusions for a net positive offset in most galaxies and a radial trend in a few of them remains valid under these two alternatives. For the cases where we originally measured a substantial positive offset ($>100$\,pc), the typical change when considering the alternative definitions is of a few times 10\% (median ${\sim}35{-}50$\%). As expected, the standard deviation is reduced when considering the weighted mean offset instead of the offset between peaks (from a mean of 1.5\,kpc to 1.0\,kpc). The galaxies that show a positive correlation between offset ($\theta$) and radius still show a positive correlation when adopting any of the other definitions of offsets, but the Spearman rank correlation coefficient changes at the level of a few multiples of 10\%. Interestingly, the derived pattern speeds are almost not affected at all by the precise definition of offsets, with a change of only a few percent.

\onecolumn

\begin{center}
\begin{longtable}{lcccccccccccc}
\caption{Impact of an alternative definition of offsets. \label{table:alternative_offsets}}\\ 
\hline\hline 
 &  \multicolumn{3}{c}{$\langle \theta_{\rm CO-H\alpha} \rangle$}   &    \multicolumn{3}{c}{$\sigma$}   &    \multicolumn{3}{c}{$\rho$}    &   \multicolumn{3}{c}{$\Omega_{\rm P}$}  \\ 
 & peak & 2$^{\rm nd}$ p.  &  weight & peak  & 2$^{\rm nd}$ p. &  weight & peak  & 2$^{\rm nd}$ p.  &  weight & peak & 2$^{\rm nd}$ p. &  weight \\ 
\hline 
    IC1954  &    206  &    176  &    137  &   1040  &   1166  &    465  &    0.29  &    0.16  &    0.18  &    32.7  &    33.6  &    28.8 \\
   NGC0628  &    236  &    220  &    154  &    948  &   1304  &    333  &    0.15  &    0.02  &    0.20  &    31.7  &    32.7  &    32.4 \\
   NGC1300  &    457  &    -90  &    426  &   3587  &   2615  &   2272  &   -0.01  &   -0.11  &   -0.05  &    23.0  &    23.1  &    23.0 \\
   NGC1385  &    265  &     54  &    189  &   1040  &   1125  &    663  &    0.31  &    0.14  &    0.27  &    26.1  &    26.4  &    25.9 \\
   NGC1512  &   -402  &   -778  &   -654  &   3083  &   3317  &   3991  &    0.19  &   -0.02  &    0.23  &    21.4  &    21.4  &    21.3 \\
   NGC1566  &    403  &    451  &    330  &   1064  &   1119  &    642  &    0.56  &    0.44  &    0.50  &    25.0  &    24.8  &    24.0 \\
   NGC1672  &    106  &    -31  &    -29  &    582  &    565  &    222  &   -0.21  &   -0.19  &   -0.41  &    15.1  &    15.0  &    14.4 \\
   NGC2090  &     33  &    124  &      1  &    596  &    600  &    138  &   -0.14  &   -0.08  &   -0.12  &   147.2  &  -218.7  &    97.9 \\
   NGC2283  &    201  &     93  &      3  &    752  &    888  &    307  &    0.32  &    0.33  &    0.24  &    28.2  &    31.9  &    34.0 \\
   NGC2566  &    737  &     73  &    -16  &   5110  &   6712  &   3456  &    0.01  &    0.01  &    0.20  &    16.1  &    16.2  &    16.2 \\
   NGC2835  &    -11  &    -10  &    -96  &    852  &   1222  &    622  &    0.21  &    0.14  &    0.18  &    29.3  &    29.3  &    29.8 \\
   NGC2997  &    250  &    358  &    494  &   2135  &   1557  &   1509  &   -0.01  &    0.04  &    0.11  &    28.2  &    28.1  &    28.0 \\
   NGC3507  &    146  &    134  &   -119  &   1757  &   2391  &    717  &    0.23  &    0.12  &   -0.21  &    27.6  &    22.8  &    23.1 \\
   NGC3627  &    176  &     90  &    201  &    962  &    984  &    371  &    0.29  &    0.34  &    0.40  &    26.6  &    29.6  &    19.1 \\
   NGC4254  &    377  &    271  &    181  &   1093  &   1086  &    603  &    0.14  &    0.04  &    0.03  &    15.2  &     2.4  &    97.0 \\
   NGC4303  &    691  &    596  &    112  &   2070  &   2393  &    663  &    0.40  &    0.13  &    0.24  &    34.8  &    35.3  &    35.3 \\
   NGC4321  &    409  &    151  &    398  &   1738  &   1243  &    537  &    0.09  &    0.08  &    0.21  &    22.5  &    22.7  &    15.7 \\
   NGC4535  &      0  &    -27  &    208  &    937  &   1030  &    793  &   -0.29  &   -0.09  &   -0.29  &    27.2  &    27.2  &    27.1 \\
   NGC4548  &    254  &     53  &     31  &    896  &    743  &    269  &   -0.03  &   -0.15  &   -0.51  &    36.7  &    36.8  &    40.8 \\
   NGC4731  &    188  &    194  &    115  &    575  &    653  &    438  &    0.23  &    0.15  &    0.38  &    25.1  &    26.0  &    24.8 \\
   NGC5194  &    399  &    155  &    456  &   1111  &   1800  &   1961  &   -0.34  &   -0.07  &    0.04  &    61.5  &    43.8  &    43.4 \\
   NGC5248  &     72  &     47  &    -66  &    798  &    679  &    476  &    0.21  &    0.20  &    0.21  &    25.3  &    25.2  &    26.1 \\
   NGC5643  &    282  &    134  &   -119  &   2114  &   2391  &    717  &    0.00  &    0.12  &   -0.21  &    35.4  &    22.8  &    23.1 \\
   NGC6744  &      1  &    134  &   -119  &   1051  &   2391  &    717  &    0.21  &    0.16  &    0.18  &    23.0  &    22.9  &    23.2 \\
\hline
\end{longtable}
\tablefoot{Mean CO-H$\alpha$ offsets, their scatter, the strength of the $\theta$--$\Omega$ correlation (Spearman rank $\rho$) and the implied pattern speed ($\Omega_{\rm P}$) for alternative definitions of offsets. `Peak' represents our nominal measurement, based on the peak emission in CO and H$\alpha$ (see Fig.~\ref{fig:method}); `2$^{\rm nd}$ peak' considers instead the azimuthal bin corresponding to the second brightest peak in CO and H$\alpha$ for each spiral arm at each radius; finally, `weighted' considers the mean azimuth weighted by CO and H$\alpha$ luminosity for each spiral at each radius.}
\end{center}

\twocolumn

\begin{figure*}[t]
\begin{center}
\includegraphics[trim=0 -200 0 0, clip,width=0.95\textwidth]{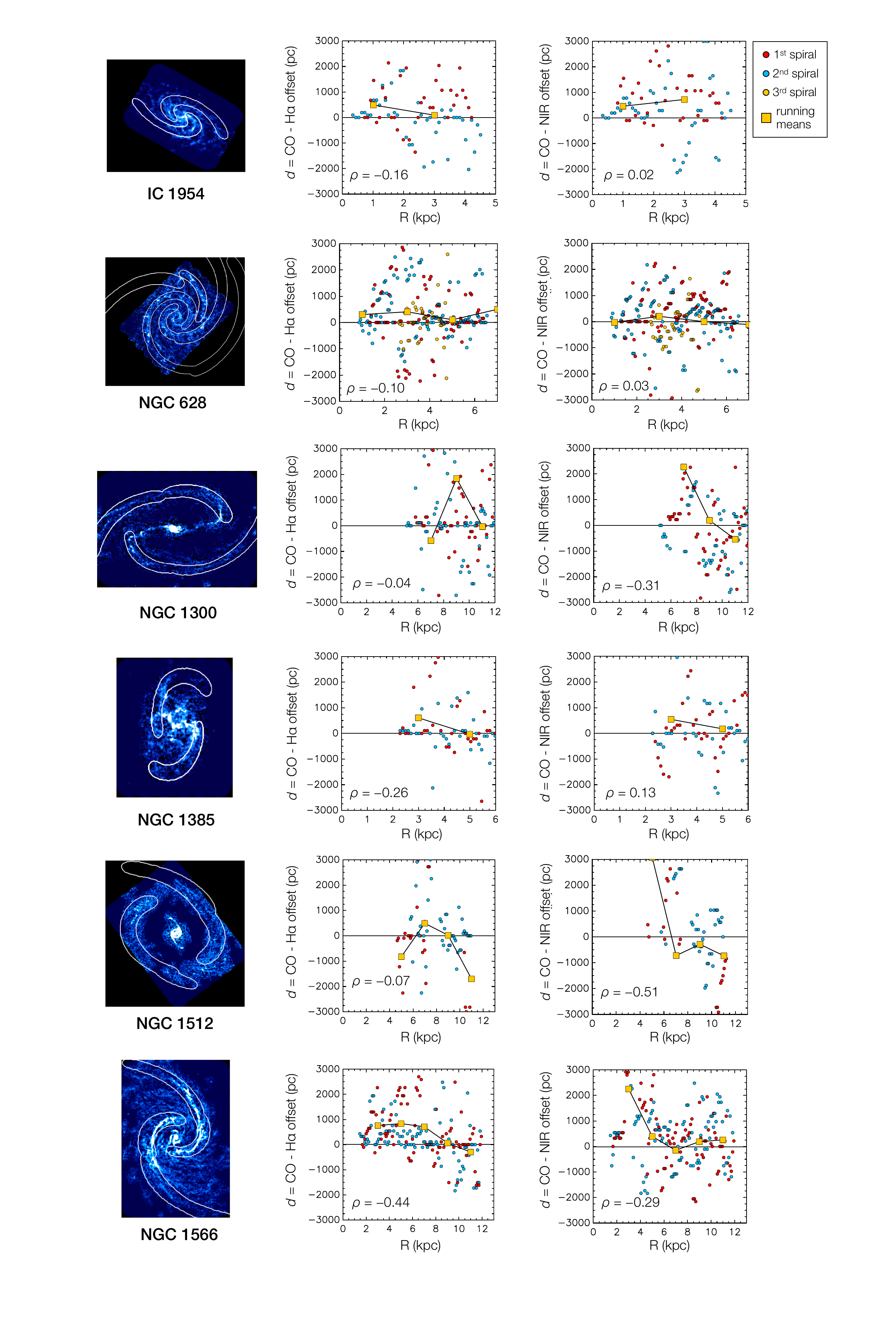}
\end{center}
\caption{Atlas of the galaxies studied in this paper showing the CO(2-1) map from ALMA with the spiral masks on top. Middle panel: CO-H$\alpha$ offsets in parsec as a function of radius. Right panel: Equivalent CO-NIR offsets. \mod{The error bars are comparable to the size of the plotted circles.} The yellow squares joined by a line show the running means in radial steps of 2\,kpc.}
\label{fig:xx}
\end{figure*}

\begin{figure*}[t]
\begin{center}
\includegraphics[trim=0 -200 0 0, clip,width=0.95\textwidth]{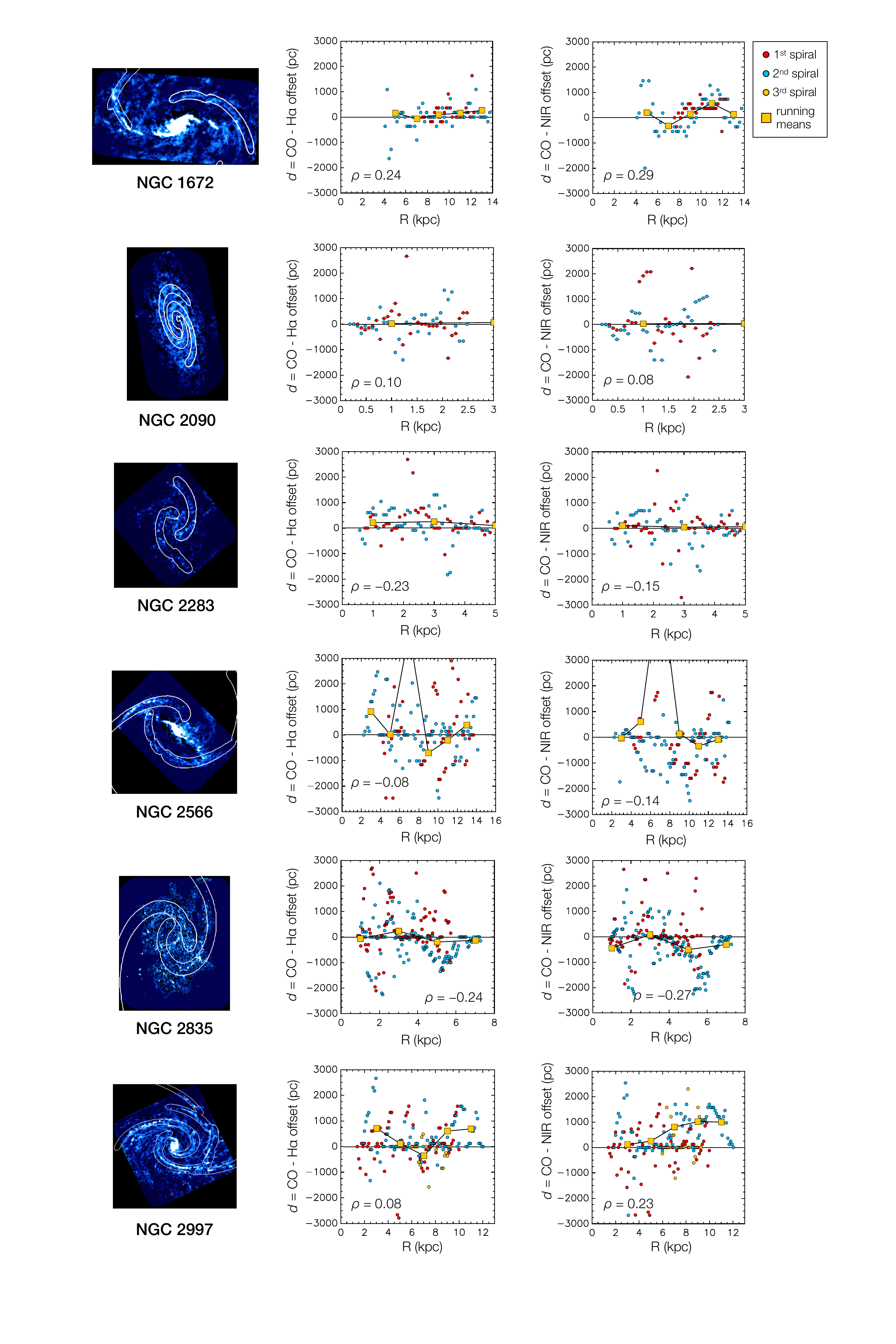}
\end{center}
\caption{Atlas of the galaxies studied in this paper (continued).}
\label{fig:atlas2}
\end{figure*}

\begin{figure*}[t]
\begin{center}
\includegraphics[trim=0 -200 0 0, clip,width=0.95\textwidth]{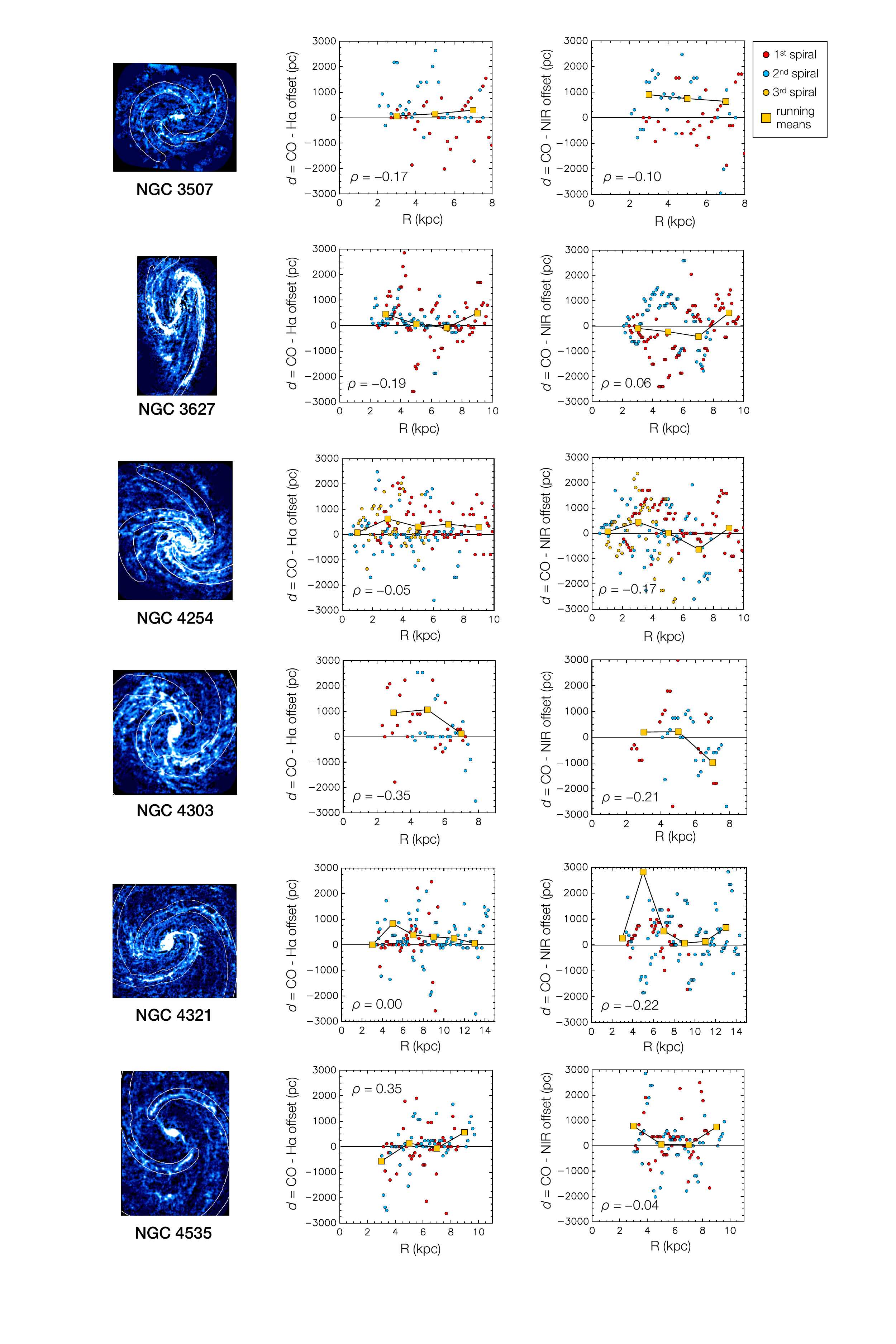}
\end{center}
\caption{Atlas of the galaxies studied in this paper (continued).}
\label{fig:atlas3}
\end{figure*}

\begin{figure*}[t]
\begin{center}
\includegraphics[trim=0 120 0 0, clip,width=0.92\textwidth]{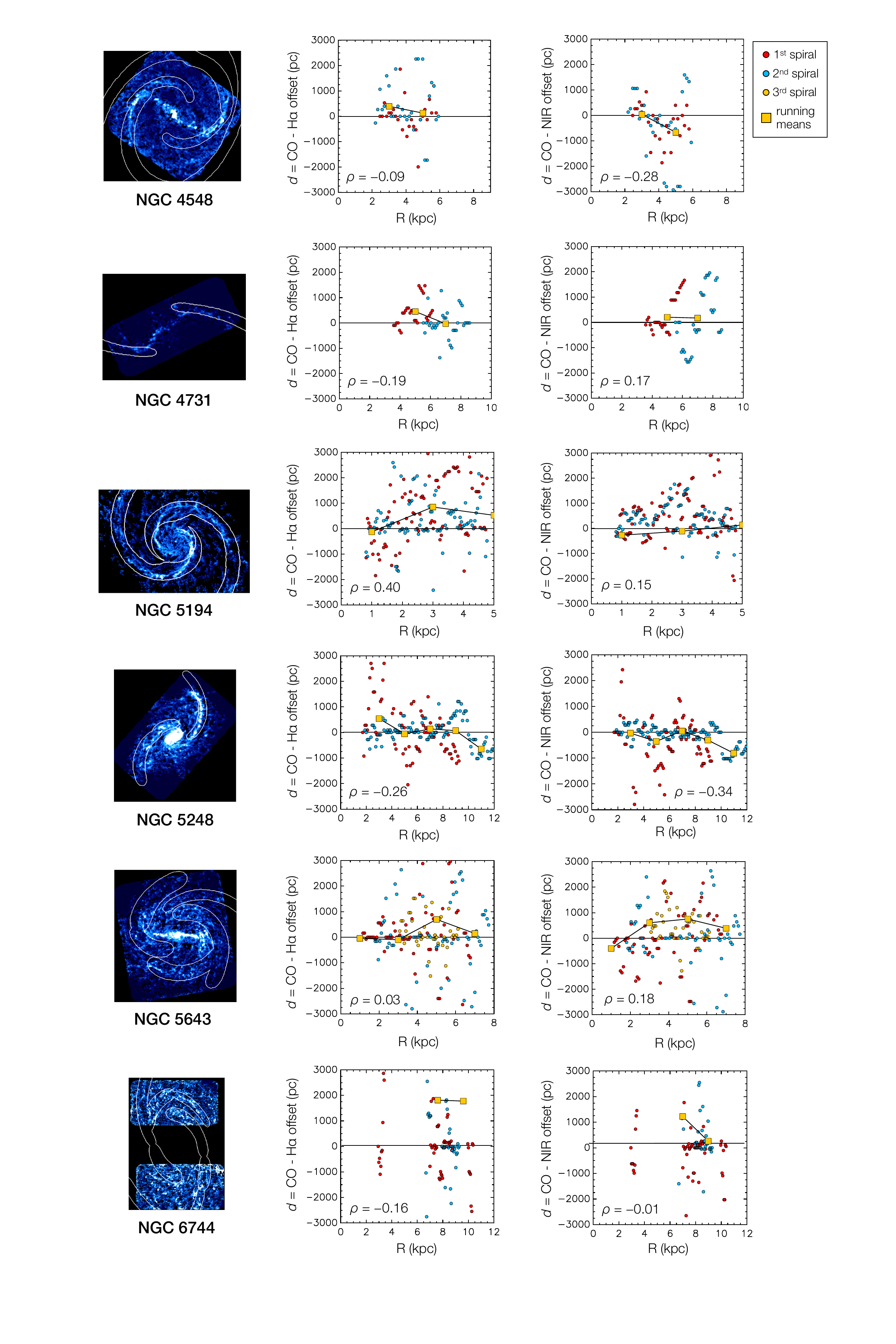}
\end{center}
\caption{Atlas of the galaxies studied in this paper, showing the CO(2-1) map from ALMA with the spiral masks on top. In the middle panel, CO-H$\alpha$ offsets in parsec as a function of radius. In the right panel, equivalent CO-NIR offsets. \mod{The error bars are comparable to the size of the plotted circles.} The yellow squares joined by a line show the running means in radial steps of 2\,kpc.}
\label{fig:atlas}
\end{figure*}

\end{appendix}

\end{document}